\documentclass[aps, prd, 12pt, superscriptaddress, amsmath, amssymb, tightenlines]{revtex4-2}
\pdfoutput=1
\usepackage{amsmath}
\usepackage{amssymb}
\usepackage[utf8]{inputenc}
\usepackage[unicode,breaklinks=true]{hyperref}
\usepackage{graphicx}
\allowdisplaybreaks[1]
\def\shortminus{\raisebox{.75pt}{-}}
\newcommand*{\defeq}{\mathrel{\vcenter{\baselineskip0.5ex \lineskiplimit0pt
                     \hbox{\scriptsize.}\hbox{\scriptsize.}}}%
                     =}

\begin{document}

\title{
A new derivation of the amplitude of asymptotic oscillatory tails of weakly delocalized
solitons
}

\author{Gyula Fodor}
\affiliation{Wigner Research Centre for Physics, 1525 Budapest 114, P.O.~Box 49, Hungary}
\author{Péter Forgács}
\affiliation{Wigner Research Centre for Physics, 1525 Budapest 114, P.O.~Box 49, Hungary}
\affiliation{Institut Denis-Poisson CNRS/UMR 7013, Université de Tours, Parc de
Grandmont, 37200 Tours, France}
\author{Muneeb Mushtaq}
\affiliation{Wigner Research Centre for Physics, 1525 Budapest 114, P.O.~Box 49, Hungary}

\date{\today}

\begin{abstract}
\setlength{\baselineskip}{1.1\baselineskip}
The computation of the amplitude, $\alpha$, of asymptotic standing wave tails of
weakly delocalized, stationary solutions in a fifth-order Korteweg-de Vries equation is revisited.
Assuming the coefficient of the fifth order derivative term, $\epsilon^2\ll1$,
a new derivation of the ``beyond all orders in $\epsilon$'' amplitude, $\alpha$, is presented. It is shown by asymptotic matching techniques, extended to higher orders in $\epsilon$, that the value of $\alpha$ can be obtained from the asymmetry at the center of the unique solution exponentially decaying in one direction.
This observation, complemented by some fundamental results of Hammersley and Mazzarino [\href{https://doi.org/10.1098/rspa.1989.0067}{Proc.~R.~Soc.~Lond.~A 424, 19 (1989)}], not only sheds new light on the computation of $\alpha$, but also greatly facilitates its numerical determination to a remarkable precision for so small values of $\epsilon$, which are beyond the capabilities of standard numerical methods.
\end{abstract}

\maketitle

\setlength{\baselineskip}{1.2\baselineskip}

\section{Introduction}
As it has been observed in numerous systems, spatially localized, ``soliton''-type excitations become ``weakly'' delocalized, by developing asymptotic ``shelves'' or oscillating tails under the influence of certain, physically important perturbations \cite{KN78, KarpMas78, KnN80}. An approximate picture of a weakly delocalized soliton would consist of a well defined core and of a small amplitude wave.
Interestingly, slowly radiating lumps in various field theories, referred to as pulsons/oscillons \cite{Bogolyubskii77, Gleiser94} belong to a related class of problems.
Even if oscillons are not solitons, they do have numerous physical applications, as many of them exhibit remarkable stability properties and extremely long lifetimes \cite{Seidel94,Copeland95,Honda2001}.
Slowly radiating oscillons have a well defined core, containing most of their energy, together with small amplitude outgoing waves.
The simple idea to approximate slowly radiating oscillons through an assumed adiabatic evolution of appropriate \textsl{stationary} configurations, called ``quasi-breathers'' -- weakly delocalized lumps with asymptotic standing wave tails \cite{Fodor06,Saffin07} has proved to be quite fruitful.
Quasi-breathers are time-periodic, and can be thought of as oscillons made stationary by incoming radiation from infinity. Determination of the amplitude of the quasi-breather wave tails is necessary to deduce the radiation rate of time-dependent oscillons.
It is rather nontrivial to compute, either numerically or perturbatively, the amplitude of quasi-breather standing wave tails, since they are exponentially suppressed, hence beyond all orders in perturbation theory.
For some recent reviews on oscillons and quasi-breathers see e.g.\ \cite{Cyncynates2021, Fodor19,Visinelli21}.

As brilliantly shown in the pioneering works of \cite{SegurKruskal91,SegurKruskal87}, and \cite{Pomeau88} the fate of spatially localized solutions in a large class of singularly perturbed equations is conveniently analyzed using methods referred to as ``beyond all orders'' asymptotics. A number of techniques have been developed, starting with asymptotic matching in the complex plane \cite{SegurKruskal91,SegurKruskal87}, to obtain an exponentially suppressed, but crucial effect of the perturbation.

The Korteweg-de Vries (KdV) equation modified by a fifth order derivative term (fKdV) has been derived in various physical contexts \cite{Kakutani69,Kawahara72,Hunter88,Benilov93}. When the fifth order derivative term
can be treated as a small perturbation, the fKdV equation admits weakly delocalized solitons with asymptotic wave tails. The amplitude of the asymptotic wave tails are typically ``beyond all orders'' and it has been computed
by several methods \cite{Pomeau88,Grimshaw95,Sun98,Boyd90,Boyd98}. In a recent paper \cite{Fodor2023} we have computed higher order perturbative corrections to the leading ``beyond all orders'' amplitude of the asymptotic wave tails, significantly extending previous result of \cite{Grimshaw95}.
In a remarkable, although somewhat lesser known work, Hammersley and Mazzarino \cite{HM}, have actually
produced \textsl{exact, spatially localized solutions} of the fKdV equation.
The localized solutions found in  Ref.\  \cite{HM}, are given in terms of convergent power series, computable to arbitrary precision.
Solutions of Ref.\ \cite{HM} are smooth on $\mathbb{R}^{+}$ with a well localized core, decreasing monotonically with exponential decay for $x\to\infty$. They are, however, regular only on the half-line, $\mathbb{R}^{+}$, and
cannot be smoothly continued to regular (decaying or even bounded) solutions to the whole real line, $\mathbb{R}$.
This latter fact also follows from the results of Refs.\ \cite{Amick91,Gunney99}, where the absence of nontrivial, globally regular decaying solutions of the fKdV equation has been shown.

The main aim of this paper is to connect two, a priori quite distinct classes of solutions of the fKdV equation :
the decaying and localized solutions of Ref.\ \cite{HM} defined on $\mathbb{R}^{+}$, on the one hand, with weakly delocalized, globally regular (and even) ones defined on $\mathbb{R}$, which are asymptotically periodic, on the other hand.

To be more precise, we relate the amplitude of asymptotic wave tails of globally regular solutions of Eq.\ \eqref{eq: HM} to the ``asymmetry'' of the localized solution's core. This asymmetry is defined as the value of the third derivative of the decaying solution on $\mathbb{R}^{+}$ at the point where its first derivative vanishes.
From our point of view, one of the main results of Ref.\ \cite{HM} is the obtention of the exact value of the
asymmetry as a function of $\epsilon$.

In fact we can connect these two different solution families only for sufficiently small values of the coefficient of the fifth order term, $\epsilon^2\ll1$.
Even if the validity of the discovered relation is limited for small values of $\epsilon$, it still seems to us not only being of interest in its own right, but also rather useful in practice, as it makes possible the computation of the amplitude of standing waves tails for arbitrarily small values of $\epsilon$. This provides a fundamentally new way for the verification of perturbative results for such small values of $\epsilon$ which have been numerically inaccessible up to now. It is very much easier to compute numerically the asymmetry of the localized solution to high precision, rather than the amplitude of the wave tail of the weakly delocalized one.
Our correspondence starts to noticeably lose both in precision and in applicability for values $\epsilon^2\gtrsim\tfrac{1}{8}$.
For such ``large'' values of the fifth order (linear) dispersion term, the fKdV approximation loses, however, its physical applicability due to other non-negligible hydrodynamical corrections, e.g.\ non-linear dispersion terms.

Clearly it is of great interest to establish analogous results in other, more complicated and/or higher dimensional cases, in particular for quasi-breathers. Perturbative results for small amplitude
quasi-breathers in various dimensions, have been found to be in reasonably good qualitative agreement with  numerical simulations \cite{Fodor09b,Fodor09c,Fodor2010}. Those results are, however, very far from the precision attained by the techniques based on asymmetry-type computations. A major problem for perturbative computations of quasi-breather tails is to go beyond the leading order term. Another difficulty is to carry out numerical simulations for sufficiently small values of the perturbative parameter in order to compare the results with the perturbative ones. A suitable generalization of the method presented in this paper is expected be quite a useful first step in that direction.

The plan of the paper is the following : In Section \ref{sec:fKdV} the transformation of the fifth order KdV equation
to the form used in Ref.\ \cite{HM} is explicited, and some of its relevant solutions are briefly discussed.
Next, in Section \ref{sec: asymm} the remarkable construction of the exact (asymmetric) solution of Ref.\ \cite{HM} is
reviewed in some detail.
In Section \ref{sec: comp} the actual computational method of the exact value of the asymmetry (valid for any value of $\epsilon>0$) is described in detail. The asymmetry, also computed by a spectral numerical code is shown to agree to the
exact results to many significant digits.
In Section \ref{sec: symm}
the perturbative construction of weakly delocalized solutions for $\epsilon\ll1$, based on Ref.\ \cite{Fodor2023},
is recapitulated. The perturbative results are exhibited up to order $\epsilon^{12}$. An improved spectral numerical method for the computation of the standing wave tails is also presented.
Section \ref{sec: rel} contains a new derivation of the amplitude of the standing wave tails of the weakly delocalized solutions for $\epsilon\ll1$. The perturbative results are shown to be in excellent agreement with very high precision
numerical calculations. Finally section \ref{sec:conc} contains our conclusions.

\section{Stationary form of fifth order KdV equation}\label{sec:fKdV}

The fifth order Korteweg-de Vries (fKdV) equation is often written as \cite{Hunter88,Pomeau88,Fodor2023}
\begin{equation}
 u_{,t} + \epsilon^2 u_{,XXXXX} + u_{,XXX}+6uu_{,X} = 0
 \,,\quad u=u(t,X)\,, \label{eq: fKdV}
\end{equation}
where $\epsilon>0$ is a (small) parameter,
$u_{,t}$ resp.\ $u_{,X}$ denote derivatives with respect to $t$ resp.\ $X$.
Considering stationary, unidirectional traveling waves of speed $c>0$, we can pass to the comoving frame, $x=X-ct$, and then solutions of Eq.\ \eqref{eq: fKdV} depending only on $x$, $u=u(x)$, satisfy the following ordinary differential equation (ODE):
\begin{equation}
 \epsilon^2u_{,5}+u_{,3}+(6u-c)u_{,1} = 0 \,,\quad \hbox{where}\quad
 u_{,n}=\frac{{\rm d}^n u}{{\rm d}x^n}\,.
 \label{eq: statfKdV5}
\end{equation}
Eq.\ \eqref{eq: statfKdV5} can be integrated once, yielding a 4th order ODE, discussed in detail in a number of papers, see e.g.\ Refs.\ \cite{HM,Hunter88,Fodor2023} :
\begin{equation}
 \epsilon^2u_{,4}+u_{,2}+3u^2-cu = M \,.
\end{equation}
By a constant shift of $u$ in Eq.\ \eqref{eq: statfKdV} one can always transform the inhomogenous term, $M$, to zero, while
$c\rightarrow\pm\sqrt{c^2+12M}$, so from now on we assume $M=0$,
\begin{equation}
 \epsilon^2u_{,4}+u_{,2}+3u^2-cu = 0\,.
 \label{eq: statfKdV}
\end{equation}
For a discussion of the physical significance of $M$, and some related subtleties see
e.g.\ Refs.\ \cite{Grimshaw95,Fodor2023}.
Next, by the following scaling transformations of Eq.\ \eqref{eq: statfKdV}; $x\rightarrow \tilde{x}/s_1$, $u\rightarrow s_2y$, with $s_1^2=c$, $s_2=\tfrac{c}{3}$,
one can transform it to the form used in Ref.\ \cite{HM}:
\begin{equation}
 \tilde{\epsilon}y_{,4}+y_{,2}+y^2-y = 0 \,,\quad
 y_{,n}=y_{,n}(\tilde{x})=\frac{{\rm d}^ny}{{\rm d}\tilde{x}^n}\,,
\label{eq: HM}
\end{equation}
where
\begin{equation}
 \tilde{\epsilon}=c\epsilon^{2} \,, \quad y=\frac{3}{c}u \,, \quad
 \tilde{x}=xc^{\frac{1}{2}} \,. \label{equytr}
\end{equation}
We note that from the relation $u(x)=\tfrac{c}{3}y(xc^{\frac{1}{2}})$,
the third derivatives, from which the asymmetry at the origin can be deduced, are related as :
\begin{equation}\label{eq: 3d}
y_{,3}(\tilde{x})=3c^{-\tfrac{5}{2}}u_{,3}(\tilde{x}c^{-\tfrac{1}{2}})\,.
\end{equation}
The celebrated (single) soliton solution of the KdV equation ($ \tilde{\epsilon}=0$ in Eq.\ \eqref{eq: HM} ),
corresponding to a travelling wave is given as
\begin{equation}\label{eq: KdVsol}
y_{\rm s}=\tfrac{3}{2}\,{\rm sech}^2(\tilde{x}/2)\,.
\end{equation}
Clearly the stationary (single) soliton, $y_{\rm s}$, is well localized in space, i.e.\ for $\tilde{x}\to\pm\infty$ $y_{\rm s}\to0$ exponentially and it has a characteristic core, where most of its mass (energy) is concentrated. Also, importantly $y_{\rm s}(\tilde{x})$ is an \textsl{even} function of $\tilde{x}$.

As already mentioned, the influence of (singular) perturbation on localized solutions often results in the appearance of ``shelves'' or ``weak delocalization''.
It has been amply discussed in the literature \cite{Amick91,Lom97,Gunney99}, that the stationary fKdV equation in the comoving frame, Eq.\ \eqref{eq: HM} for $\tilde{\epsilon}>0$, does not admit nontrivial bounded solutions on the real line, $\mathbb{R}$, with decaying boundary conditions for $\tilde{x}\to\pm\infty$. Bounded solutions on $\mathbb{R}$, for sufficiently small values of $\tilde{\epsilon}>0$, tend to \textsl{asymptotically periodic functions} (oscillating tails) for $\tilde{x}\to\pm\infty$, \cite{Hunter88,SegurKruskal87,Pomeau88,Boyd90,Benilov93,Grimshaw95,Boyd98}.

\begin{figure}[!hbt]
 \centering
 \includegraphics[width=110mm]{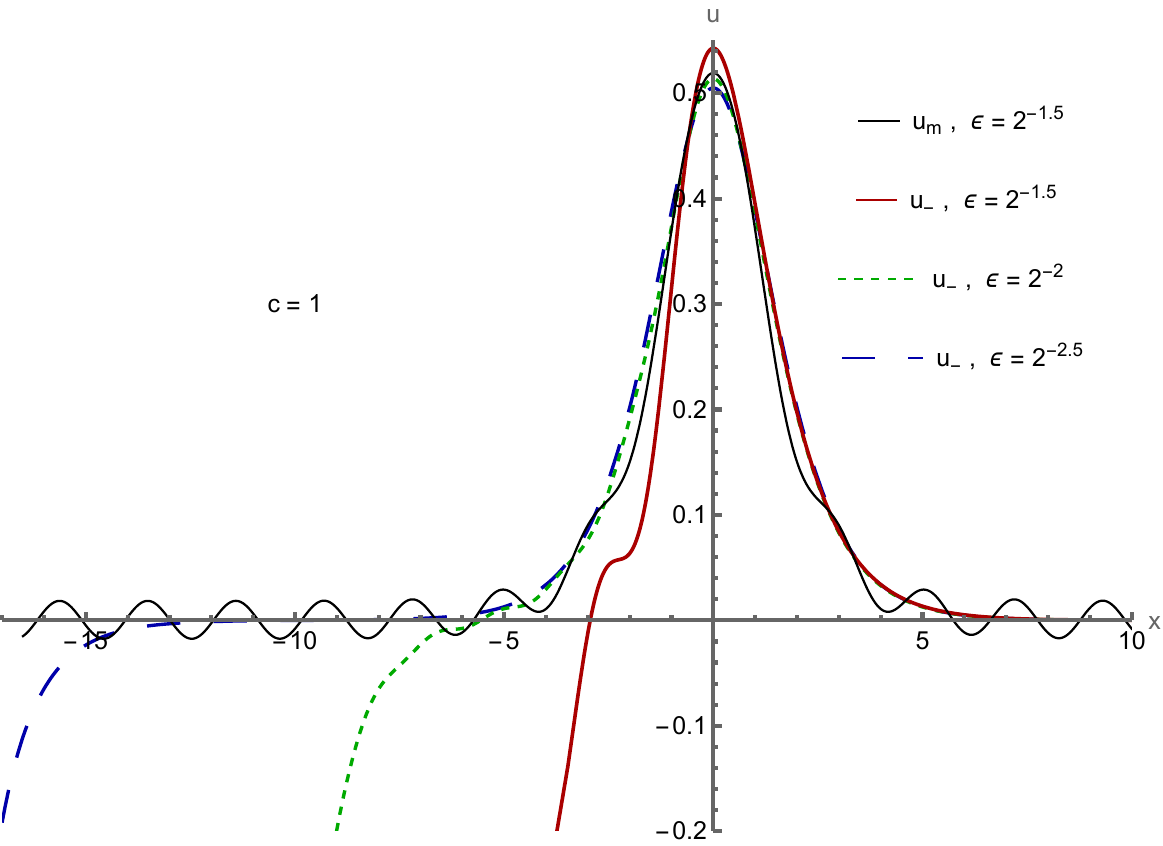}
\caption{\label{figexpcmpnum} The black curve shows the minimal tail symmetric solution $u_{\rm m}$ of Eq.~\eqref{eq: statfKdV} for the parameter values $\epsilon=2^{-1.5}$, $c=1$. For the same values of $(\epsilon, c)$, the red curve corresponds to the asymmetric solution, $u_{-}$, decaying to zero for $x>0$, and running into a singularity on $\mathbb{R}^{-}$. The singularity moves farther from the center as $\epsilon$ decreases, as illustrated by the green and the blue curves. Since the tail-amplitude decreases very fast with decreasing $\epsilon$, the symmetric solutions are not plotted for the smaller $\epsilon$ values.}
\end{figure}

On the other hand, Hammersley and Mazzarino \cite{HM}, have explicitly constructed smooth solutions, $y(\tilde{x})$, of Eq.\ \eqref{eq: HM} for any $\tilde{\epsilon}>0$ satisfying one-sided decay conditions, say for $\tilde{x}\to\infty$.
The solution of Ref.\ \cite{HM} is unique for $\tilde{\epsilon}$ fixed, it decreases monotonically for $\tilde{x}>0$ with a maximum at the origin, $\tilde{x}=0$, defined by the vanishing of the first derivative, $y_{,1}(0)=0$.
Importantly these solutions are not even functions of $\tilde{x}$, in that $y_{,3}(0)\ne0$.
As already noted above, in view of the non-existence theorems of Refs.\ \cite{Amick91,Gunney99}, regular solutions decaying for $\tilde{x}\to\infty$ of Eq.\ \eqref{eq: HM} on $\mathbb{R}^{+}$ cannot be continued to globally regular ones on $\mathbb{R}$. We remark here, that according to our numerical experiments asymmetric solutions of Eq.\ \eqref{eq: statfKdV}, decaying on $\mathbb{R}^{+}$ run generically into a moving pole-type singularity on $\mathbb{R}^{-}$. Figure \ref{figexpcmpnum}.\ depicts some numerically calculated symmetric and asymmetric solutions of Eq.~\eqref{eq: statfKdV}.

The results of Ref.\ \cite{HM} make it possible to obtain a simple, quantitative measure of asymmetry, defined by the value of $y_{,3}(0)$, of regular, decaying solutions as a function of $\tilde{\epsilon}>0$. In the following we shall connect the value of $y_{,3}(0)$ of decaying solutions on $\mathbb{R}^{+}$ to the amplitude of the asymptotic wave tails of bounded, globally regular solutions of Eq.\ \eqref{eq: HM} which are \textsl{even functions} on $\mathbb{R}$.

\section{Construction of asymptotically decaying, asymmetric solutions}\label{sec: asymm}

In this section we outline the main steps of the ingenious construction of Hammersley and Mazzarino \cite{HM}, for further details, the reader should consult the original paper.

Statement of the problem : find regular, monotonically decreasing solutions on $\mathbb{R}^{+}$ of Eq.\ \eqref{eq: HM}, subject to the boundary conditions :
\begin{equation}\label{eq: bc1}
y_{,1}(0)=0\,,\quad \lim_{\tilde{x}\to\infty}y(\tilde{x})=0\,,\quad \hbox{and}\quad y_{,1}(\tilde{x})<0\ \; \forall \tilde{x}>0\,.
\end{equation}
According to Ref.\ \cite{HM} existence and uniqueness of solutions with bounded derivatives, $y_{,n}(\tilde{x})$, for $\forall n\geq1$, satisfying \eqref{eq: bc1}, follows from adopting the arguments given in Ref.\ \cite{HM2}.
Eq.\ \eqref{eq: HM} admits a familiar first integral :
\begin{equation}\label{eq: 3thO}
\tilde{\epsilon}(y_{,1}y_{,3}-\tfrac{1}{2}y_{,2}^2)+\tfrac{1}{2}y_{,1}^2
=\tfrac{1}{2}y^2-\tfrac{1}{3}y^3\,,
\end{equation}
where the constant of integration has been chosen to be zero, since by assumption $y$ and its derivatives tend to zero for $\tilde{x}\to\infty$.
Because Eq.\ \eqref{eq: 3thO} is an autonomous 3rd order ordinary differential equation (ODE),
one can reduce it to a 2nd order ODE by
considering $y_{,n}$ to be a function of $y$.
As $y(\tilde{x})$ is a monotonically decreasing function for $\tilde{x}\geq 0$ by assumption, both its inverse function, $\tilde{x}(y)$, and $y_{,1}(y)$ are well defined on the interval $(0,y(0))$. We introduce a new independent variable, $z\propto y$ as
\begin{equation}\label{eq: transf0}
y=Yz\,,
\quad Y=y(0)\,.
\end{equation}
The new variable, $z$, varies in the interval $z\in(0,1)$, because
$\tilde{x}\to\infty$ corresponds to $z\to0$.
This tacitly presupposes $Y=y(0)>0$, which has been shown to hold for $\forall\tilde{\epsilon}>0$ in Ref.\ \cite{HM}.
Introducing a new unknown, $f(z)$, as
\begin{equation}\label{eq: transf1}
y_{,1}=-Yf^{\tfrac{1}{2}}\,,
\end{equation}
one finds
\begin{equation}\label{eq: transf}
y_{,2}=\tfrac{1}{2}Yf_{,1}\,,\quad y_{,3}=-\tfrac{1}{2}Yf^{\tfrac{1}{2}}f_{,2}\,,
\quad f_{,n}=\frac{{\rm d}^nf}{{\rm d}{z}^n}\,,
\end{equation}
so that Eq.\ \eqref{eq: 3thO} is transformed to the 2nd order ODE :
\begin{equation}\label{eq: 2ndO}
\tilde{\epsilon}(ff_{,2}-\tfrac{1}{4}f_{,1}^2)+f = z^2-\tfrac{2}{3}Yz^3\,.
\end{equation}
 The solution, $f(z)$, is conveniently represented by the power series :
\begin{equation}\label{eq: fser}
f(z)=\lambda^2z^2\left(1-\sum_{n=1}^{\infty} c_nz^n\right)\,,
\end{equation}
where the coefficients, $c_n$, are unrelated to the speed, denoted by $c$ in the previous Section, and $\lambda$ is the (unique) positive root of the quartic
\begin{equation}\label{eq: lamda}
\tilde{\epsilon}\lambda^4+\lambda^2 -1=0\,,\quad \lambda^2=\frac{\sqrt{1+4\tilde{\epsilon}}-1}{2\tilde{\epsilon}}\,.
\end{equation}
Choosing the positive root of Eq.\ \eqref{eq: lamda} ensures that for $\tilde{x}\to\infty$ the decay condition, $y(\tilde{x})\sim {\rm{e}}^{-\lambda \tilde{x}}$, holds. From Eq.\ \eqref{eq: 2ndO} one obtains the following recurrence relations for the coefficients $c_n$~:
\begin{equation}\label{eq: c-recurr}
c_1=\tfrac{2}{3}Y/(1+4\eta^{-1})\,,
\quad q_nc_n=\frac{1}{2}\sum_{r=1}^{n-1}p_{nr}c_r c_{n-r}\,,\quad n\geq2\,,
\end{equation}
where
\begin{equation}\label{eq: pq}
q_n=\frac{(n+1)^2+\eta}{(n+1)^2+1}\,,\quad p_{nr}=1-\frac{5}{2}\frac{r(n-r)}{(n+1)^2+1}\,,\quad \eta=\frac{1}{\tilde{\epsilon}\lambda^4}\,.
\end{equation}
We also note that $\eta=\eta(\tilde{\epsilon})$ can be expressed in terms of $\tilde{\epsilon}$ as:
\begin{equation}\label{eq: eta}
\eta(\tilde{\epsilon})= 1+\tfrac{1}{2}{\tilde{\epsilon}}^{-1}\left( 1 + \sqrt{1+4{\tilde{\epsilon}}}\right)\,.
\end{equation}
In order to satisfy the boundary condition $y_{,1}(0)=0$, the series \eqref{eq: fser} should converge for $z\to1$ and
$f(1)=0$. That these conditions can be satisfied, is not at all obvious. It has been shown in Ref.\ \cite{HM} that for any $\tilde{\epsilon}>0$ it is possible to find a $c_1>0$ such that
\begin{equation}\label{eq: c-eq}
\sum_{n=1}^{\infty}c_n=1\,.
\end{equation}
The proof of the validity of Eq.\ \eqref{eq: c-eq} is a crucial point of Ref.\ \cite{HM}.
If $c_1>0$, from Eqs.\ \eqref{eq: c-recurr} and \eqref{eq: pq} one obtains that all other $c_n>0$. Then from Eq.\ \eqref{eq: c-eq} it follows
that $0<c_n<1$ for $\forall n\geq 1$, which in turn implies that $f(z)$ is an analytic function of $z$ for $|z|< 1$, and $f(1)=0$. It is also shown in Ref.\ \cite{HM} that the radius of convergence of the series \eqref{eq: fser} is exactly $1$. Also $f(z)>0$ for $z\in(0,1)$, and the boundary conditions \eqref{eq: bc1} are satisfied, indeed. Moreover $Y$ satisfies the inequality $\tfrac{3}{2}<Y<\tfrac{3}{2}(1+4\eta^{-1})$, which ensures non-triviality and boundedness of the solutions for $\forall \tilde{\epsilon}>0$.
Clearly, once $c_1$ has been found, the decaying solution in $\mathbb{R}^{+}$, $y(\tilde{x})$,  is completely determined, albeit implicitly.

It remains to compute the asymmetry $y_{,3}(0)$. The difficulty of this last step is due to the non-analyticity of $f(z)$, which has a branch point at $z=1$.
We recall that the derivatives of $y$ at $\tilde{x}=0$, denoted by
\begin{equation}
 Y_n=y_{,n}(0) \ ,
\end{equation}
are all finite for $n\geq1$.
Therefore from Eqs.\ \eqref{eq: transf} one finds for the limit $z\to1$
\begin{equation}\label{eq: y2}
 Y_2=\tfrac{1}{2}Yf_{,1}(1)=-\tfrac{1}{2}\lambda^2Y\sum_{n=1}^{\infty}nc_n\,,\quad  Y_3=-\tfrac{1}{2}Y
 \lim_{z\to1}f^{\tfrac{1}{2}}f_{,2}\,.
\end{equation}
Since $Y_2$ is finite and $Y>0$, from Eq.\ \eqref{eq: y2} one sees that for $z\to1$, $f(z)\sim (z-1)f_{,1}(1)$, from where one can deduce
\begin{equation}\label{eq: f2}
f_{,2}\sim -Y_3(-\tfrac{1}{2}YY_2)^{-\tfrac{1}{2}}(1-z)^{-\tfrac{1}{2}}\,,\quad{\rm for}\quad z\to1\,,
\end{equation}
implying also that unless $Y_3=0$, the sum, $\sum_{n=1}^{\infty}(n+2)(n+1)c_nz^{n}$, diverges for $z\to1$, necessitating more refined methods to express the asymmetry, $Y_3$, in terms of the coefficients $c_n$.
In Ref.\ \cite{HM} a positive lower bound is established for $Y_3$ :
\begin{equation}\label{eq: boundy3}
Y_3\geq2^{-15} {\tilde{\epsilon}}^{\,\shortminus\tfrac{3}{4}}\eta^{\tfrac{7}{4}}
\exp({-\pi{\tilde{\epsilon}}^{\,{\shortminus}\tfrac{1}{2}}})\,,
\end{equation}
which result captures, remarkably, both the correct ${\tilde{\epsilon}}\to0$ limiting behaviour of $Y_3$,
$Y_3\sim {\tilde{\epsilon}}^{\,\shortminus\tfrac{5}{2}}\exp({-\pi{\tilde{\epsilon}}^{\,{\shortminus}\tfrac{1}{2}}})$,
and its large, ${\tilde{\epsilon}}\to\infty$ limit,
$Y_3\sim {\tilde{\epsilon}}^{\,\shortminus\tfrac{3}{4}}$.

Using the analicity of $f(z)$ inside the unit disk in the complex $z$ plane, Hammersley and Mazzarino find that $Y_3$ is given as
\begin{equation}\label{eq: y3}
Y_3=\lambda^2(-\tfrac{1}{2}\pi YY_2)^{\tfrac{1}{2}}\,\mathbb{L}\,,\quad
\mathbb{L}=\lim_{n\to\infty}n^{\tfrac{5}{2}}c_n\,,
\end{equation}
where
\begin{equation}\label{eq: yy2}
 Y_2=-\lambda^2Y\sqrt{c_1(4+\eta)-\eta}\,,\quad\hbox{and}\quad Y=\frac{3}{2}c_1(1+4\eta^{-1})\,.
\end{equation}
We note here a slight disagreement between Eq.\ \eqref{eq: y3} and Eq.\ (4.10) of Ref.\ \cite{HM}, due to a missing factor of $1/2$ from Eq.\ (4.10).

We would like to point out that the result of Eq.\ \eqref{eq: y3} can also be deduced, from what is often referred to as transfer theorems in analytical combinatorics \cite{FlajOdl90}. Such transfer theorems relate the asymptotic behaviour of the coefficients, $c_n$, for $n\to\infty$ of a generating function, analytic inside the unit disk, to its singularities at $|z|=1$. In our case
we have $f_{,2}(z)\sim (1-z)^{-\tfrac{1}{2}}$ for $z\to1$, and a simple application of the transfer theorem of Ref.\ \cite{FlajOdl90} yields immediately  the asymptotic behaviour of the coefficients, $c_n$,
\begin{equation}\label{eq: cnasymp}
\lim_{n\to\infty} c_n \sim n^{-\tfrac{5}{2}}\,,
\end{equation}
and then one easily reproduces Eq.\ \eqref{eq: y3}.

In summary, all regular, asymptotically decaying solutions on $\mathbb{R}^{+}$ of Eq.\ \eqref{eq: 2ndO} are given by Eq.\ \eqref{eq: fser}, in terms of convergent series, where all the coefficients, $c_n$, for $n\geq2$ are determined, once for some fixed value of ${\tilde{\epsilon}}$, $c_1$ has been found.
The value of $c_1$ is determined by the somewhat implicit equation Eq.\ \eqref{eq: c-eq}, where all terms $c_n$, for $n\geq2$ are to be expressed in terms of $c_1$. Moreover Eq.\ \eqref{eq: c-eq}
contains unbounded powers of the unknown, $c_1$, therefore it is something of a challenge to obtain precise enough values of $c_1$. In addition, the actual computation of the asymmetry, $y_{3}(0)=Y_3$, also necessitates the determination of the quantity $\mathbb{L}$ which is not that straightforward. The difficulty to obtain reliable numerical estimates for the value of $\mathbb{L}$ is due to the slow convergence of its determining sequence, \eqref{eq: y3}, for large values of $n$. The smaller ${\tilde{\epsilon}}$ is, the slower is the convergence.

\section{Computational issues}\label{sec: comp}

In this Section we outline our computational procedures to determine solutions of Eq.\ \eqref{eq: 2ndO} in terms of $c_n$, which differ somewhat from those employed in Ref.\ \cite{HM}. Instead of using some approximation scheme to solve
Eq.\ \eqref{eq: c-eq} for $c_1$, we obtain approximations to it from the large $n$ asymptotic behaviour of the coefficients, $c_n$, themselves.
This might sound paradoxical at first, since it is precisely the unknown $c_1$ which determines all the $c_n$ for $n\geq2$.
An important point is to exploit the homogeneity of the recurrence relation, Eq.\ \eqref{eq: c-recurr} for the coefficients $c_n$ :
\begin{equation}\label{eq: cn-scaling}
c_1\to c_1(s)\defeq sc_1\,,\quad c_n\to c_n(s)\defeq s^nc_n\,,
\end{equation}
which permits one to choose $c_1(s)=s$ and view then Eq.\ \eqref{eq: c-eq} as an equation for $s$.
We display as an illustration, the first three non-trivial coefficients $c_n(s)$ :
\begin{equation}\label{eq: c2-c4}
c_2(s) =\frac{15 s^2}{4 (\eta +9)}\,,\quad c_3(s) =\frac{45 s^3}{(\eta +9) (\eta +16)}\,,\quad
c_4(s)=\frac{315 (6 \eta +59) s^4}{2(\eta +9)^2 (\eta +16) (\eta +25)}\,.
\end{equation}
As shown in Ref.\ \cite{HM}, the sequence of the unique positive roots, $s_N$, $N=1\,,2...$ of the correspondingly truncated polynomial equations
\begin{equation}\label{eq: c-trunc}
\sum_{n=1}^{N}c_n(s)=1\,,
\end{equation}
satisfy
\begin{equation}\label{eq: s-solns}
1=s_1>s_2>s_3>\ldots>s_N>\ldots>\frac{1}{2}\,,
\end{equation}
therefore the decreasing sequence $\{s_N\}_{N=1\,,2\ldots}$
converges clearly to a positive limit, $s_\infty\geq1/2$.
Furthermore, an important result established in Ref.\ \cite{HM} is that the coefficients defined by $\lim_{N\to\infty} c_n(s_{N})=
c_n(s_{\infty})\defeq c_n$ do satisfy Eq.\ \eqref{eq: c-eq}.
To compute approximate values of $s_{\infty}$ for a given $\eta(\tilde{\epsilon})$, Ref.\ \cite{HM} used
the obvious approximation, i.e.\ to solve numerically Eq.\ \eqref{eq: c-trunc} for $s_N>0$ with $N$ large enough and use $s_N\approx s_{\infty}$.

In the present work we have employed a different method to obtain approximations for $s_{\infty}$ to rather high precision. Instead of solving Eq.\ \eqref{eq: c-trunc} numerically with $N$ sufficiently large for
$s_N\approx s_{\infty}$, we solve (numerically) the recurrence relation Eq.\ \eqref{eq: c-recurr} for the quantities $c_n(s=1)=c_n(1)$, starting with $c_1(1)=1$.
Computing then the ratio $r_n=c_n(1)/c_{n+1}(1)$ for $n$ sufficiently large,
we obtain an approximation for $s_{\infty}$, i.e.\ $r_n\approx s_{\infty}$. The reason that the ratio, $r_n$, can serve as an approximant for $s_{\infty}$, follows directly from Eqs.\ \eqref{eq: y3}-\eqref{eq: cnasymp} describing the asymptotic behaviour of $c_n=c_n(s_{\infty})$ :
\begin{equation}\label{eq: c-as}
c_n\to\mathbb{L}n^{-\tfrac{5}{2}}\left(1+{\cal{O}}\left(\frac{1}{n}\right)\right)\,,\quad n\to\infty\,,
\end{equation}
recalling that $c_n=c_n(1)s_{\infty}^n$, it immediately follows that for $n\to\infty$ :
\begin{equation}\label{eq: r-as}
\frac{c_n}{c_{n+1}}=\frac{c_n(1)}{c_{n+1}(1)s_{\infty}}\to 1+ {\cal{O}}\left(\frac{1}{n}\right)\,,\quad {\rm i.e.}\quad
r_n\to s_{\infty}\left( 1+ {\cal{O}}\left(\frac{1}{n}\right)\right)\,.
\end{equation}
The speed of convergence of $r_n$ is considerably slowed down by a series of finite $n$ corrections of the type ${\cal{O}}(1/n^m)$, $m=1\,,2\,,\ldots\,$.
A closer inspection of the asymptotic, $n\to\infty$, behaviour of $c_n$, together with the application of transfer theorems, reveals that all finite $n$ corrections to the ratio $r_n$ decay as \textsl{integer} powers of $1/n$. Therefore this is a textbook case to increase the rate of convergence of the sequence $\{r_n\}$ considerably, by making use of higher order Richardson extrapolation.
Let us recall a compact form of the $m$-th order Richardson extrapolation used in our case \cite{BenderOrszag}:
\begin{equation}\label{eq: Rich}
R_m(r_n)=\sum_{k=0}^{m}r_{n+k}\frac{(-1)^{m+k}(n+k)^m}{k!\,(m-k)!}\,.
\end{equation}
For example, for $\tilde{\epsilon}=1/8$, in standard double precision ($\sim$ 16 decimal digits), one obtains already 8 decimal digits correctly of $s_{\infty}=0.772875581\ldots$, from not more than $27$ coefficients, $c_n(1)$, by $m=7$th order Richardson extrapolation. We note that in Ref.\ \cite{HM} 4000 terms and a first order Richardson extrapolation has been used to obtain 6 decimal digits of $c_1$ and of the asymmetry, $Y_3$, for this value of $\tilde{\epsilon}$.
As $\tilde{\epsilon}$ decreases, the number of necessary coefficients, $c_n(s)$, increases.
For the values, $2^{-25}\leq\tilde{\epsilon}\leq2^{-10}$, considered in Ref.\ \cite{HM}, to obtain 6 significant decimal digits for the asymmetry, 20000 of the coefficients $c_n(s)$, has been computed using quadruple precision, combined with a first order Richardson extrapolation.
As it should be clear, the order of Richardson extrapolations is strongly limited by the available numerical precision of $r_n$, since the number of subtractions induced by the alternating signs in Eq.\ \eqref{eq: Rich} increases with the order of the extrapolation. Therefore, it is necessary to increase the numerical precision beyond that of the standard one, using suitable software (in this paper ARB \cite{arbweb,Johansson17} and Wolfram Mathematica), permitting to use relatively high order ($\sim20-30$) Richardson extrapolations.
For the smallest values of $2^{-30}\leq\tilde{\epsilon}\leq2^{-25}$ considered in this paper,
we have found that performing the computations with the arbitrary precision software ARB to 48 decimal digits,
from not more than 200 $c_n$ and Richardson extrapolations of order $m\leq30$, the value of the asymmetry is obtained
already to 6 significant decimal digits. To increase the number of significant digits of the asymmetry to 12 it has been found sufficient to increase the number of decimal digits in ARB to 60.

The optimal order of the Richardson extrapolation can be found by searching for the minimal value of the difference $\Delta R_{m,n}=R_m(r_n)-R_{m-1}(r_n)$, choosing some fixed starting term $r_n$ and increasing the extrapolation order one by one from $m=0$. We note that the most time consuming task is the computation of the coefficients, $c_n(1)$, for large values ($n\gtrsim1000$) of $n$, while computing the Richardson extrapolants, $R_m(r_n)$, goes very fast. Usually, it is a more efficient strategy to increase the order of the Richardson extrapolation together with the number of computational digits, instead of significantly increasing the number of computed coefficients, $c_n(1)$, which permits to decrease the order of Richardson extrapolations. According to our experience, in most cases, the difference $\Delta R_{m,n}$ provides a reasonable error estimate for the result.

The calculation of the limit, $\mathbb{L}$, in Eq.\ \eqref{eq: y3} is also greatly improved by the use of Richardson extrapolation due to the same, $1/n^m$-type, corrections encountered when approximating the value of $s_\infty$.
As pointed out in Ref.\ \cite{HM}, for values of $\tilde{\epsilon}\lesssim0.01$, it is necessary
to further accelerate the convergence of the sequence $\{n^{5/2}c_n\}$, to avoid the computation of an exceedingly large number of terms.
The transformation, $c_n\Rightarrow v_n$, put forward in Ref.\ \cite{HM}
\begin{equation}\label{eq: vn}
v_n=n^{\tfrac{5}{2}}c_n\,q_0\,q_1\dots q_n\,,\quad Q\defeq\prod_{n=0}^\infty q_n =\frac{\sinh(\pi{\sqrt{\eta}})}{{\sqrt{\eta}}\sinh(\pi)}\,,
\end{equation}
where $q_n$ is defined in Eq.\ \eqref{eq: pq}, yields an alternative expression for $\mathbb{L}\,$:
\begin{equation}\label{eq: vnlim}
\mathbb{L}=\frac{1}{Q}\lim_{n\to\infty}v_n \,.
\end{equation}
This approach has also been used in this paper, together with sufficiently high-order Richardson extrapolations.

We now summarize the algorithm used for the calculation of the asymmetry. We choose some $\tilde\epsilon$ value
and carry out all calculations to a given number of digits precision. High precision floating point number calculations can be performed either by some algebraic manipulation software (Mathematica, Maple, SageMath, etc.) or by using free open source packages in C language (CLN \cite{clnweb}, ARB \cite{arbweb}). Starting from $c_1(1)=1$ we compute the $c_n(1)$ up to some order $n=N$ from the recurrence relation \eqref{eq: c-recurr}. The necessary computational time can be halved exploiting the symmetry $r\to n-r$ of $p_{nr}$. Then Richardson extrapolation is used
on $r_n=c_n(1)/c_{n+1}(1)$ to approximate the value of $s_\infty$ from Eq.\ \eqref{eq: r-as}.
The second derivative at the center, $Y_2$, is found from Eq.\ \eqref{eq: yy2}. The value of $\mathbb{L}$ is approximated by Richardson extrapolation of the series $v_n$, defined in Eq.\ \eqref{eq: vn}.
Finally, the asymmetry, $Y_3$, can be obtained from Eq.\ \eqref{eq: y3}.
Applying the above method we have reproduced all the six digits of the results in Table $1$ of Ref.~\cite{HM}.
For example, choosing $\tilde\epsilon=2^{-12}$ and computing the coefficients $c_n(1)$ up to $n=200$ with $56$ decimal digits precision, for the computation of $Y_3$ the optimal order Richardson extrapolation turns out to be $15$. This way the asymmetry, $Y_3\approx9.20154645\cdot10^{-77}$ is obtained up to $19$ decimal digits of precision.
The computational methods presented in this paper allow for the determination of $c_1$ and $Y_3$ up to $1000$ (or even more) decimal digits of precision.

As Supplemental Material \cite{supp1} a Mathematica notebook and an equivalent C language file that uses the ARB library \cite{arbweb} are provided, to carry out computations of $c_1$ and $Y_3$ to high precision for a wide range of $\tilde\epsilon$.

The asymptotically decaying asymmetric solution of Eq.~\eqref{eq: HM} can also be calculated directly by using a high precision spectral numerical method. The semi-infinite interval $0\leq\tilde x\leq\infty$ can be transformed into the $\hat x\in[-1,1]$ region by introducing the new independent variable
\begin{equation}
 \hat x=\frac{\tilde x-\xi}{\tilde x+\xi} \ ,
\end{equation}
where $\xi>0$ is a constant representing a length scale. The monotonically decreasing $y(\hat x)$ function remains smooth under this transformation, and can be searched for by usual Chebyshev polynomial expansion.
It is sufficient to impose the vanishing of ${{\rm d}y}/{{\rm d}\hat{x}}$ at $\hat x=-1$ corresponding to the center,
no further boundary conditions are necessary.
The reason for this is that the compactification method excludes both oscillatory and unbounded solutions.
This phenomenon is referred to as ``behavioral boundary conditions'' in Ref.\ \cite{Boyd13}.
This compactification approach would not work for the calculation of the symmetric soluton because of the infinitely many oscillations in the tail region.
For $\tilde\epsilon\lesssim 2^{-8}$ the central third derivative that we intend to calculate, becomes so small that a very large number of collocation points and more than $16$ decimal digits precision becomes necessary. Since in the present paper the numerical construction of the decaying solution is just for consistency checking, we do not go here into more details. The method we applied is very similar to the one described in our previous paper \cite{Fodor2023}. The obtained results for $y_{,3}(0)$ agree to many decimal digits with the computationally much less expensive Hammersley-Mazzarino method described earlier.
For illustration we note that our highest resolution numerical computations used $2200$ collocation points with $540$ decimal digits of precision (applying the ARB library \cite{arbweb}). The running times took about $7$ hours on a desktop computer, providing an absolute error for $y_{,3}(0)$ as tiny as $10^{-397}$, whose magnitude becomes comparable to the value of the asymmetry for $\tilde\epsilon\approx 2^{-17}$. This type of spectral numerical calculations would be too slow and memory-hungry for $\tilde\epsilon\lesssim 2^{-17}$, when the third derivative becomes smaller than this error value.
The numerical evaluation of the exact solution does not necessitate significant computational resources,
and it is applicable for any value of $\tilde\epsilon$.
We expect that the asymmetry computations based on perturbative expansions and spectral numerical methods will remain applicable for more complicated analogous problems, such as the radiation of scalar field oscillons, where it is unlikely that analogous results to that of Hammersley and Mazzarino could be found.

\section{Globally regular, non-decaying symmetric solutions}\label{sec: symm}

In the preceding sections, following Ref.\ \cite{HM}, we have presented
the exact solution of Eq.\ \eqref{eq: HM} decaying exponentially for $\tilde{x}\to\infty$ and being smooth on $\mathbb{R}^+$.
As already mentioned, in view of the results of Refs.\ \cite{Amick91,Gunney99}
this solution cannot be smoothly extended to the whole real line, $\mathbb{R}$, with decaying boundary conditions for $\tilde{x}\to-\infty$.
Our numerical experiments indicate that solutions decaying for $\tilde{x}\to\infty$, become unbounded when
continued for negative values of $\tilde{x}$ as $\sim-840\tilde{\epsilon}/(\tilde{x}+x_0)^4$, where $x_0$ is determined by the value of $\tilde{\epsilon}$, cf.\ Fig.\ \ref{figexpcmpnum}.

As already mentioned, the existence of globally regular solutions of Eq.\ \eqref{eq: statfKdV} on $\mathbb{R}$,
asymptoting some periodic function for $x\to\pm\infty$ has been established for sufficiently small values of $\epsilon$, \cite{Hunter88}.
In Refs.\ \cite{Amick92,Sun98} the existence of a one-parameter family of globally regular solutions of
Eq.\ \eqref{eq: statfKdV} for sufficiently small values of $\epsilon>0$, has been proven.
Assuming that for $|x|\to\infty$ the solutions are well approximated by those of the \textsl{linearized version} of Eq.\ \eqref{eq: statfKdV}, one can seek them in the form $\sim {\rm{e}}^{ikx/\epsilon}$. One finds that $k$ satisfies
\begin{equation}\label{eq: k}
k^4-k^2-\underbrace{{\epsilon}^2c}_{\displaystyle{\tilde{\epsilon}}}=0\,,\quad \hbox{with real roots}\quad k^2=\tfrac{1}{2}(1+\sqrt{1+4\tilde{\epsilon}})\,.
\end{equation}
In the following $k$ shall denote the positive root of Eq.\ \eqref{eq: k}.
The real roots in Eq.\ \eqref{eq: k}, determine the standing wave tail of the symmetric (even with respect to reflections $x\to-x$) solution. This tail can be written as $\sim\alpha\sin(k{|x|}/{\epsilon}-\delta)$, where $\alpha>0$, $\delta$ are constants, $k/{\epsilon}>0$ is the wave-number. The tail amplitude, $\alpha$, turns out to be exponentially small in ${\epsilon}$, and $\delta$
corresponds to the asymptotic phase.
Physically, a particular member of this family is singled out, namely the one where the amplitude of the tail is \textsl{minimal}, $\alpha=\alpha_{\rm m}$ for some value of $\delta=\delta_{\rm m}$. Both $\alpha_{\rm m}$ and $\delta_{\rm m}$ are functions of the parameter $\epsilon$.

The two imaginary roots of Eq.\ \eqref{eq: k} correspond to an exponentially growing and a decaying mode, which we write
following \cite{Grimshaw95}, as $\sim {\rm{e}}^{\pm 2\gamma x}$ with $\gamma>0$.
The real and imaginary roots of Eq.\ \eqref{eq: k} are related as:
\begin{equation}\label{eq: kgamma}
4(\epsilon\gamma)^2+1=k^2\,.
\end{equation}
The wave speed, $c$, can be expressed  as:
\begin{equation}\label{eq: cgamma}
c=4\gamma^2(1+4\gamma^2\epsilon^2)=4\gamma^2k^2\,.
\end{equation}
Following previous numerical works \cite{Boyd91,Boyd95}, in Ref.\ \cite{Fodor2023} a multiple precision pseudo-spectral numerical code has been developed to construct globally regular solutions of Eq.\ \eqref{eq: statfKdV}, aimed to
obtain the ${\epsilon}$ dependence of the minimal tail amplitude, $\alpha_{\rm m}$, to high precision. The numerically constructed (symmetric) solutions of Eq.\ \eqref{eq: statfKdV} have been found to be in excellent agreement with the analytical ones, obtained by perturbation theory in ${\epsilon}$ and by asymptotic matching.

We now briefly recapitulate the perturbative construction of weakly delocalized solutions of Eq.\ \eqref{eq: statfKdV} assuming $0<\epsilon\ll1$, for details see Ref.\ \cite{Fodor2023}. We also follow closely the notations of Ref.\ \cite{Fodor2023}.
We recall that weakly delocalized solitons have a well defined, fast decaying core
with an exponentially suppressed asymptotic standing wave tail. The core part, $u_{\rm c}$, is obtained
by expanding the solution, $u$, of Eq.\ \eqref{eq: statfKdV} in a power series in ${\epsilon}$ around the KdV soliton, i.e.\ writing :
\begin{equation}
 u_{\rm c}^{\scriptscriptstyle(N)}=\sum_{n=0}^{N}u_n{{\epsilon}}^{2n} \,, \quad u_0=2\gamma^2{\rm sech}^2(\gamma{x})\,,\label{eq: core-exp1}
\end{equation}
where $N$ determines the order of the approximation to $u_{\rm c}$.
In Refs.\ \cite{Grimshaw95,Fodor2023}, the wave-speed, $c$ \eqref{eq: cgamma}, has been considered as a function of $\epsilon$, while $\gamma$ has been assumed to be an $\epsilon$ independent constant. Note that in the present paper another choice, $c=1$, has been made for all numerical computations.
According to Ref.\ \cite{Fodor2023} $u_n$ can be expressed as
\begin{equation}
 u_n=\gamma^{2n+2}\sum_{j=1}^{n+1}u_{n,j}\,\mathrm{sech}^{2j}(\gamma x)
\,,\label{eq: core-exp2}
\end{equation}
where $u_{n,j}$ are (rational) numbers. The localized, core-part of the solution, $u_{\rm c}$, is defined as the (formal) expansion \eqref{eq: core-exp1} for $N\to\infty$, it is to be understood as an asymptotic series in $\epsilon$.
This approximation has an optimal order, $N_{\epsilon}$ determined by $\epsilon$, which for small enough values of $\epsilon$ behaves as
$N_{\epsilon}\sim 1.56/(\epsilon\sqrt{c})$.

In Ref.~\cite{Fodor2023} we have used a spectral numerical method to find solutions of Eq.~\eqref{eq: statfKdV} that are symmetric with respect to $x=0$. For smaller $\epsilon$ parameters this procedure can be made significantly more efficient by first calculating the optimally truncated approximation $u_{\rm c}^{\scriptscriptstyle(N_\epsilon)}$, which is correct up to ${\cal O}(\epsilon^{2N_\epsilon+2})$ terms, and then numerically solving the equation for the difference $v=u-u_{\rm c}^{\scriptscriptstyle(N_\epsilon)}$,
\begin{align}
 &\epsilon^2v_{,4}+v_{,2}+3v^2-c v +6u_{\rm c}^{\scriptscriptstyle(N_\epsilon)}v
 = R_\epsilon \ , \label{eq: optdiff} \\
 & R_\epsilon=-\epsilon^2u_{\mathrm{c}\,,4}^{\scriptscriptstyle(N_\epsilon)}
 -u_{\mathrm{c}\,,2}^{\scriptscriptstyle(N_\epsilon)}
 -3\left(u_{\rm c}^{\scriptscriptstyle(N_\epsilon)}\right)^2
 +c u_{\rm c}^{\scriptscriptstyle(N_\epsilon)} \ .
\end{align}
The approximation $u_{\rm c}^{\scriptscriptstyle(N_\epsilon)}$ and its residual $R_\epsilon$ can be calculated very quickly to many digits of precision using the algorithm presented in \cite{Fodor2023}. The function $u_{\rm c}^{\scriptscriptstyle(N_\epsilon)}$ decays exponentially as $x\to\infty$. The residual $R_\epsilon$ is very small for small $\epsilon$, since $u_{\rm c}^{\scriptscriptstyle(N_\epsilon)}$ is an ideal approximation to the real solution. The function $R_\epsilon$ has about $N_\epsilon$ oscillations with exponentially decreasing amplitude, then the exponential decrease continues monotonically. The same spectral numerical method can be applied as earlier in \cite{Fodor2023} to solve the differential equation \eqref{eq: optdiff} in a finite interval with matching to the linearized tail at the outer boundary. The current approach is more efficient, because the function $v$ remains as small as the oscillating tail even in the core region (see Fig.~2 of \cite{Fodor2023}). However, because of the numerous oscillations in $v$, it is still necessary to use large number of collocation points and extended precision (more than $16$ decimal digits) computations. Our C language code uses the ARB library for arbitrary-precision ball arithmetic\cite{arbweb}. The obtained numerical results for the minimal tail amplitude are presented in Table \ref{tablerhoalp}.
\begin{table}[!hbtp]
 \centering
 \begin{tabular}{|c||c|c|c|c|}
  \hline
  $\epsilon$ & $\alpha_{\rm m}$ & $\delta_{\rm m}$ & $N_\epsilon$
  & $u_{-,3}(0)$ \\
  \hline
  \hline
  $2^{-1}$ & $4.8\cdot10^{-2}$ & $1.2$ & $1$
  & $7.10760589389\cdot10^{-1}$ \\
  \hline
  $2^{-2}$ & $1.53\cdot10^{-3}$ & $0.749$ & $4$
  & $8.11499816225\cdot10^{-2}$ \\
  \hline
  $2^{-3}$ & $3.2525301\cdot10^{-8}$ & $0.37253582$ & $11$
  & $1.62487857959\cdot10^{-5}$ \\
  \hline
  $2^{-4}$ & $1.94383743304\cdot10^{-18}$ & $0.187147747984$ & $23$
  & $7.92099610003\cdot10^{-15}$ \\
  \hline
  $2^{-5}$ & $1.27243717968\cdot10^{-39}$ & $0.0937046561766$ & $49$
  & $4.16436578363\cdot10^{-35}$ \\
  \hline
  $2^{-6}$ & $1.17039547447\cdot10^{-82}$ & $0.0468692914439$ & $99$
  & $3.06718215011\cdot10^{-77}$ \\
  \hline
  $2^{-7}$ & $2.29646599329\cdot10^{-169}$ & $0.0234367851656$ & $199$
  & $4.81567051778\cdot10^{-163}$ \\
  \hline
  $2^{-8}$ & $2.12961253593\cdot10^{-343}$ & $0.0117186606062$ & $400$
  & $3.57282878973\cdot10^{-336}$ \\
  \hline
  $2^{-9}$ & $4.49453680829\cdot10^{-692}$ & $0.00585936382454$ & $802$
  & $6.03243642172\cdot10^{-684}$ \\
  \hline
 \end{tabular}
 \caption{Numerically calculated values of the minimal tail amplitude $\alpha_{\rm m}$, the corresponding phase $\delta_{\rm m}$, and the order of optimal truncation $N_\epsilon$ for the case $c=1$. The last column gives the central third derivative of the asymmetric solution $u_{-}$.}
 \label{tablerhoalp}
\end{table}
The result for a given $\epsilon$ contains an error of order $\alpha_{\rm m}^2$, since we match to a linearized tail solution at the outer boundary. For $\epsilon\geq 2^{-3}$ only the reliable digits are presented, while for smaller $\epsilon$ we give only the first $12$ decimal digits of the results. In order to get more precise amplitudes for $\epsilon\geq 2^{-3}$ we would need to match to higher amplitude nonlinear spatially periodic solutions, which is beyond the scope of the present paper. In the last column of Table \ref{tablerhoalp} we list the central third derivative of the asymmetric right decaying solution, $u_{-,3}(0)$, calculated by the analytical methods in Sec.~\ref{sec: comp}. The relation between $\alpha_{\rm m}$ and $u_{-,3}(0)$ will be the subject of the next section.

The delocalized, standing wave tail part of the symmetric solution has been constructed in the WKB approximation, see e.g.\ Ref.\ \cite{BenderOrszag}, in the following way:
We look for approximate solutions of Eq.\ \eqref{eq: statfKdV} as $u=u_{\rm c}+u_w$ neglecting the nonlinear term. Then $u_w$ is readily found to obey the equation:
\begin{equation}
 \epsilon^2u_{w,4}+u_{w,2}+6u_{\rm c} u_w-cu_w = 0 \,,\quad u_{w,n}=\frac{{\rm d}^n u_w}{{\rm d}x^n}\,.\label{eq: uw}
\end{equation}
Solutions of Eq.\ \eqref{eq: uw} tending to $\alpha\sin(k{|x|}/{\epsilon}-\delta)$ for $|x|\to\infty$, are expected to provide ${\cal O}(\alpha)$ approximations to those of Eq.\ \eqref{eq: statfKdV}.
In practice one also has to chose some finite order approximation in $\epsilon$ for $u_{\rm c}$, but since $u_w$ is beyond all orders in $\epsilon$ this does not cause any order-mixing problems.
In Ref.\ \cite{Fodor2023} high order WKB approximations of the solution of Eq.\ \eqref{eq: uw}, $u_w^{\scriptscriptstyle{\rm WKB}}$, have been obtained, which can be formally written as :
\begin{equation} \label{eq: uwsol}
u_w^{\scriptscriptstyle{\rm WKB}} = \beta\,\exp\left[{\sum_{n=1}^{\infty} A_{2n}\epsilon^{2n}}\right]\sin\left(\frac{kx}{\epsilon}-\delta_w-\delta(x)\right)
\end{equation}
where $\beta$, $\delta_w$ are real constants, while the phase-function, $\delta(x)$, is written as :
\begin{equation}
\delta(x)=\sum_{n=0}^{\infty} \tilde{A}_{2n+1}\epsilon^{2n+1}\,.
\end{equation}
The phase-function, $\delta(x)$, is to be distinguished from the asymptotic phase, denoted as $\delta$ previously and in \cite{Fodor2023}.
The expansion terms, $\tilde{A}_k$, have been determined to rather high $\sim 100$ order using algebraic manipulation codes.
The even index (amplitude) coefficients, $A_{2n}$ are even functions in $x$, they tend to zero exponentially fast for $x\to\pm\infty$, e.g.\ the first one is given as
$A_2=15\gamma^2{\rm sech}^{2}(\gamma{x})$. The odd index (phase) functions, $\tilde{A}_{2n+1}$ are odd in $x$, they tend to constants for $x\to\pm\infty$. For example the first one is given as $\tilde{A}_1=6\gamma\tanh(\gamma{x})$. The asymptotic behaviour of the phase function, $\delta(x)$, has been written in \cite{Fodor2023} as
\begin{equation}\label{eq:phase_exp}
\lim_{x\to\infty}\delta(x)=\delta_{\rm m}=\sum_{n=0}^{\infty}\tilde{\delta}_{2n+1}\epsilon^{2n+1}\,,\quad {\rm e.g.}\ \tilde{\delta}_1=6\gamma\,.
\end{equation}
The determination of the amplitude of the approximate WKB-solution, $\beta$, amounts to solving the notoriously difficult connection problem. Following the pioneering works of \cite{SegurKruskal91,SegurKruskal87}, the connection problem has been solved by an asymptotic matching procedure in the complex plane to leading order in $\epsilon$ in Ref.\ \cite{Pomeau88}.
This work has been extended in Refs.\ \cite{Grimshaw95,Sun98} where it has been pointed out that there are more general solutions depending on a phase parameter corresponding to $\delta_w$. As argued
in \cite{Fodor2023}, the amplitude, $\alpha$, of the asymptotic tail of globally regular (symmetric) solutions of \eqref{eq: statfKdV} with asymptotic phase $\delta$ are not generally minimal, their amplitude being determined as $\alpha=\alpha_{\rm m}/\cos(\delta-\delta_{\rm m}) + {\cal O}(\alpha_{\rm m}^2)$.

Within the framework of the perturbative asymptotic matching there is a unique, minimal tail amplitude solution, $u_{\rm m}(x)$, with $\alpha=\alpha_{\rm m}(\epsilon)$  corresponding to $\delta_w=0$, which has the asymptotic phase $\delta=\delta_{\rm m}$.
The result up to ${\cal{O}}(\epsilon^6)$ for the \textsl{minimal amplitude} wave-tail has been presented in Ref.\ \cite{Fodor2023}. We would like to point out, that we have considerably extended the result in \cite{Fodor2023} by working out $\alpha_{\rm m}$ to higher orders in $\epsilon$. The presentation of the details is beyond the scope of this paper, but we show some of the obtained results. The order $2N$ approximation to the minimal amplitude $\alpha_{\rm m}$ can be given as:
\begin{equation}\label{eq: alpha}
 \alpha_{\rm m}^{(N)}=\frac{\mathcal{K}}{\epsilon^2}
 {\rm e}^{-\tfrac{\pi k}{2\gamma\epsilon}}
  \left( 1-\sum_{n=1}^{N}\xi_{n}(\gamma\epsilon)^{2n}
  \right)\,,\quad
  \mathcal{K}\approx19.96894735876096\!\cdot\!\pi \,,
\end{equation}
where the wave-number is $k=(1+4\gamma^2\epsilon^2)^{1/2}$.
The first six coefficients, $\xi_n$, are given as
\begin{align}\label{eq: xis}
\notag
\xi_1&=5\,,& \xi_2&\approx6.54406819358\,,  &\xi_3&\approx474.413839489\,,\\
\xi_4&\approx4233.41235937\,, &\xi_{5}&\approx111053.952710\,, &\xi_{6}&\approx1782156.51421\,.
\notag
\end{align}
For convenience we give the expansion of $k=k(\epsilon)$ up to ${\cal{O}}(\epsilon^{2N})$ :
\begin{equation}\label{eq: k-exp}
 \exp\left(-\frac{k\pi}{2\gamma\epsilon}\right)=
 \exp\left(-\frac{\pi}{2\gamma\epsilon}\right)
 \left[ 1-\pi\gamma\epsilon+\frac{\pi^2}{2}\gamma^2\epsilon^2
 -\left(\frac{\pi^2}{6}-1\right)\pi\gamma^3\epsilon^3
 +\ldots\right] \,.
\end{equation}
Using \eqref{eq: k-exp} it is straightforward to obtain the fully expanded result for $ \alpha_{\rm m}^{(N)}$.\ To avoid unnecessarily long expressions, in the following we keep $k$ in the exponential term of Eq.\ \eqref{eq: alpha}.

\begin{figure}[!hbt]
 \centering
 \includegraphics[width=110mm]{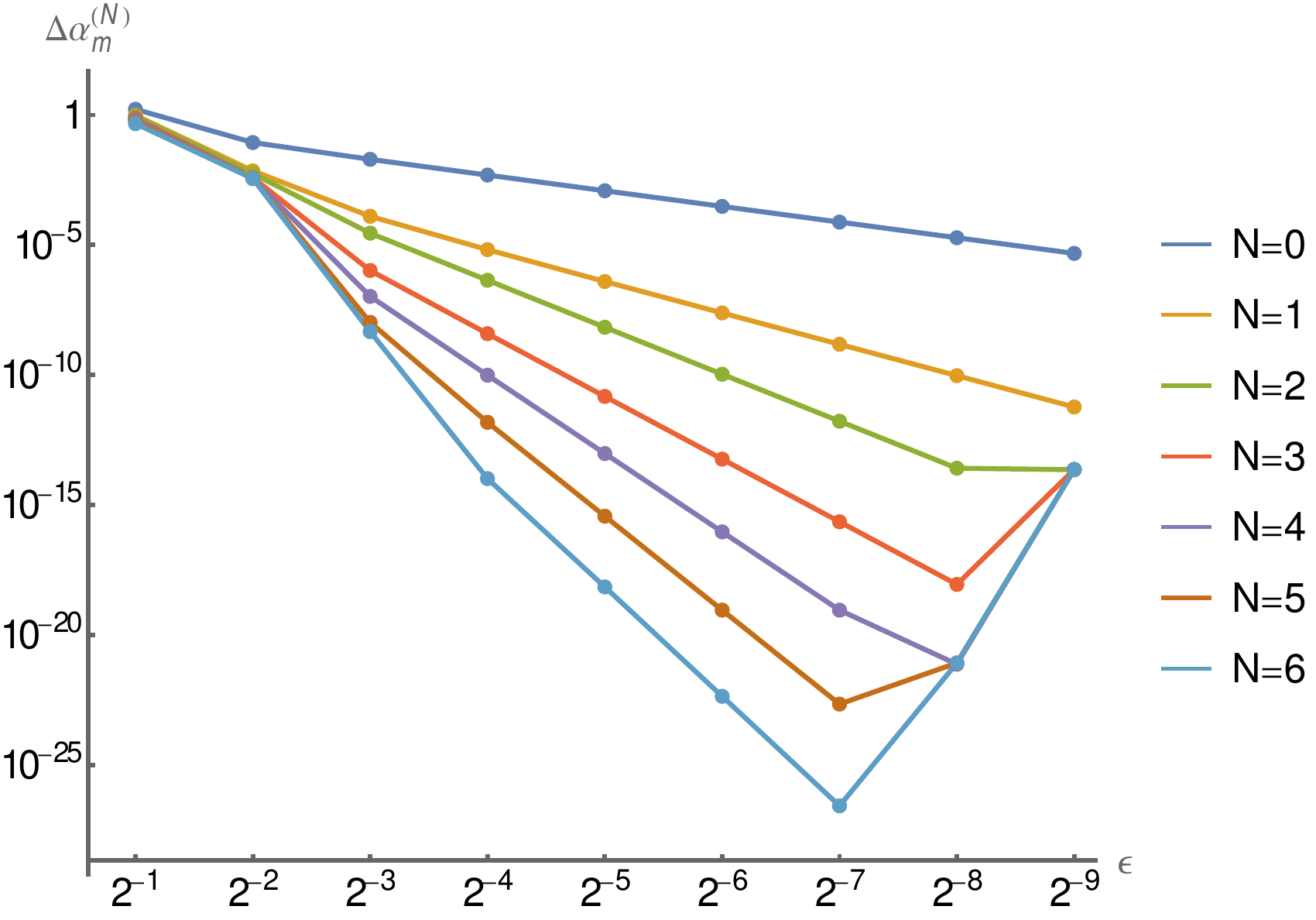}
\caption{\label{figasymsym} Relative difference
$\Delta\alpha_{\rm m}^{(N)}=\left|(\alpha_{\rm m}-\alpha_{\rm m}^{(N)})/\alpha_{\rm m}\right|$ of the numerical results and the $N$-th order expansion results as a function of $\epsilon$.}
\end{figure}

The coefficients $\mathcal{K}$ and $\xi_n$ can be obtained to hundreds of decimal digits of precision relatively easily, using the methods presented in \cite{Fodor2023}, and applying Richardson extrapolation. The high precision results are useful for consistency checking when comparing them with the numerical results in Table \ref{tablerhoalp}, and also
with the analytical results in the following sections.
It is to be emphasized that beside higher order $\epsilon^{2M}$, ($M>N+1$) corrections to $\alpha_{\rm m}^{(N)}$ in
Eq.\ \eqref{eq: alpha}, there are also other corrections of the type $\exp\{-{m\pi k}/{(2\gamma\epsilon)}\}$, with $m>1$ integer,
which are, however, exponentially suppressed with respect to $m=1$.
In Fig.~\ref{figasymsym} we plot the relative difference of the numerical results $\alpha_{\rm m}$ and perturbative results for $\alpha_{\rm m}^{(N)}$ up to order $N=6$ for the values of $\epsilon$ in Table \ref{tablerhoalp}.
From Fig.~\ref{figasymsym} it is apparent that for ``large'' values of $\epsilon>2^{-2}$ the expansion results for 
$\alpha_{\rm m}^{(N)}$ are not particularly meaningful. 
On the other hand, the numerical results obtained from our spectral code become less and less reliable for $\epsilon<2^{-7}$, since in those cases we cannot have enough collocation points to describe well the large number of oscillations in $u$. An important conclusion one can draw from Figure \ref{figasymsym} is that the error of the $N$-th order approximation, $\alpha_{\rm m}^{(N)}$, decreases as $\epsilon^{2N+2}$, moreover, its relative error is well estimated by the next order term, $\xi_{N+1}(\gamma\epsilon)^{2N+2}$. It is quite interesting, that the calculated coefficients, $\xi_{n}$, do not seem to increase factorially with $n$, which is generally expected to occur for asymptotic series. The points depicted on Fig.~\ref{figasymsym} are consistent with convergent behaviour for the depicted $\epsilon$ range. It is tempting to conjecture that the sum in \eqref{eq: alpha} is convergent for $\epsilon\lesssim1/2$, but the clarification of this would clearly require further analysis.

\section{Relating asymmetric and symmetric (even) solutions}\label{sec: rel}

We shall now relate two important type of solutions of Eq.\ \eqref{eq: statfKdV} discussed in Sections \ref{sec: asymm} and \ref{sec: symm} (see Fig.~\ref{figexpcmpnum}).
According to Ref.\ \cite{HM} the ``asymmetric'', decaying solution of Sec.\ \ref{sec: asymm}, $u_{-}$, has been proven to be unique. The minimal tail amplitude solution,
$u_{\rm m}$, approximated in Sec.\ \ref{sec: symm}, is also expected to be unique among the globally regular, ``symmetric'' ones, (unfortunately we are not aware of a formal uniqueness proof).
For sufficiently small values of $\epsilon$, the amplitude of the asymptotic tail of $u_{\rm m}$ is many orders of magnitude smaller than that of its core. Therefore the heuristic description of
$u_{\rm m}$ as the linear superposition of a fast decaying core and of a small amplitude standing wave tail
is expected to be a very good approximation.

\begin{figure}[!hbt]
 \centering
 \includegraphics[width=110mm]{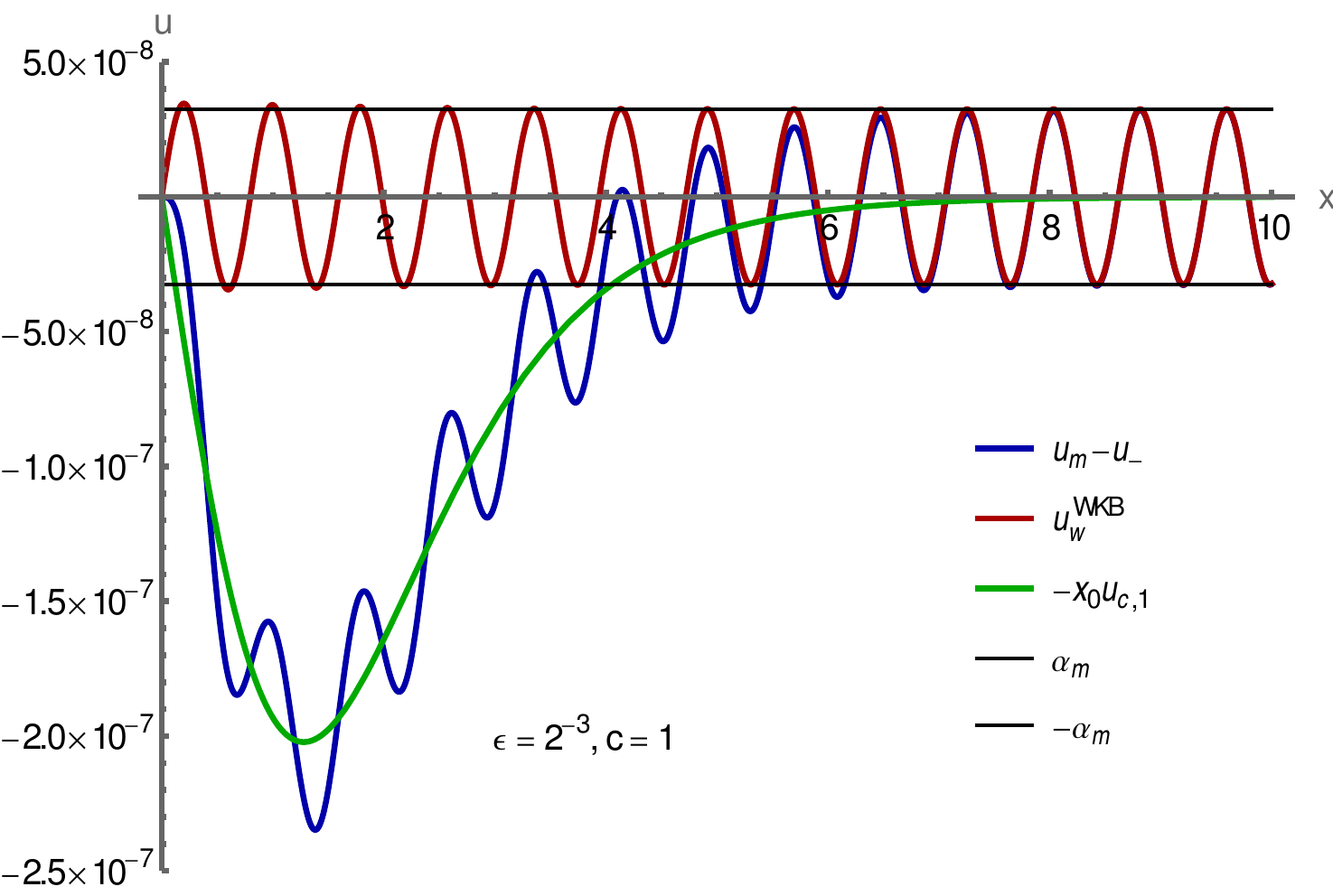}
\caption{\label{figudiff} The blue curve shows the difference $u_{\rm m}-u_{-}$ for $\epsilon=2^{-3}$ and $c=1$. The green curve, $- x_0u_{{\rm c},1}$, is a translational mode of the linearized equation \eqref{eq: uw}, with $x_0=-1.03371\cdot 10^{-6}$. The difference of these two yields the red curve, $u_w^{\scriptscriptstyle{\rm WKB}}\approx u_{\rm m}-u_{-}+x_0u_{{\rm c},1}\approx u_{\rm m}(x)-u_{-}(x-x_0)$, where the equalities hold up to error of $4\cdot 10^{-13}$ for the selected $\epsilon$ value.}
\end{figure}

Consider the difference $\Delta=u_{\rm m}-u_{-}$, which tends exponentially fast for $x\gg1/\epsilon$ to the tail part of $u_{\rm m}$.
Our numerical simulations clearly show that $\Delta$ is of order $\alpha_{\rm m}$ for $\forall x\in\mathbb{R}^{+}$, i.e.\ \textsl{also in the core region}. Since both $u_{\rm m}$ and $u_{-}$ satisfy Eq.\ \eqref{eq: statfKdV} it is easily seen that $\Delta$ satisfies Eq.\ \eqref{eq: uw} with $u_w$ replaced by $\Delta$, when omitting the quadratic term.
Therefore we approximate $\Delta$ by a suitable solution of \eqref{eq: uw}, and write the minimal tail solution as
\begin{equation}
 u_{\rm m}\approx u_{-}+u_w^{\scriptscriptstyle{\rm WKB}} - x_0u_{{\rm c},1}\,,
\quad \mathrm{for} \quad \beta=\alpha_{\rm m} \,, \ \delta_w=0 \,,
\label{eq: apprum}
\end{equation}
where $x_0$ is a constant to be determined.
The term $u_{{\rm c},1}$ is the translational zero mode of the (formal) solution, $u_{\rm c}$, of Eq.\ \eqref{eq: statfKdV} hence also that of Eq.\ \eqref{eq: uw}.
The necessity to include the translational mode, $u_{{\rm c},1}$, in Eq.\ \eqref{eq: apprum} can be understood
from the fact that the approximation of the minimal amplitude tail, $u_{w}^{\scriptscriptstyle{\rm WKB}}$
is \textsl{an odd function} with respect to the origin, while $u_{\rm m}$ is even and $u_{-}$ has been defined to obey the boundary condition, $u_{-,1}(0)=0$.
An alternative viewpoint for Eq.\ \eqref{eq: apprum} is that one can cancel the tail of $u_{\rm m}$ by subtracting a linear perturbation approximated by $u_w^{\scriptscriptstyle{\rm WKB}}$, and then obtain a shifted version of the asymmetric solution $u_{-}$, so that its maximum is at $x=x_0$,
\begin{equation}
 u_{\rm m}(x)\approx u_{-}(x-x_0)+u_w^{\scriptscriptstyle{\rm WKB}}(x) \,.
\end{equation}
In Fig.~\ref{figudiff} we plot the difference of the symmetric solution $u_{\rm m}$ and asymmetric solution $u_{-}$ for the parameter choice $\epsilon=2^{-3}$, $c=1$.

Taking derivatives of the approximation \eqref{eq: apprum} at $x=0$ one obtains the following relations:
\begin{align}
&u_{\rm m}(0)=u_{-}(0)\,,\label{eq:um0}\\
&{u_{w,1}^{\scriptscriptstyle{\rm WKB}}(0)}
- x_0{u_{{\rm c},2}(0)}=0\,,\label{eq:um1}\\
&u_{{\rm m},2}(0)=u_{-,2}(0)\,,\label{eq:um2}\\
&u_{-,3}(0)+u_{w,3}^{\scriptscriptstyle{\rm WKB}}(0) - x_0u_{{\rm c},4}(0)=0\,,
\label{eq:um3}
\end{align}
using that $u_{w}^{\scriptscriptstyle{\rm WKB}}(0)=u_{w,2}^{\scriptscriptstyle{\rm WKB}}(0)=0$ for $\delta_w=0$ and $u_{{\rm c},1}(0)=u_{{\rm c},3}(0)=0$. Eqs.\ \eqref{eq:um0}-\eqref{eq:um3} are expected to be valid up to arbirary orders in $\epsilon$ and to first order in ${\cal O}(\alpha_{\rm m})$.

As detailed in \cite{Fodor2023} the ${\cal O}(\alpha_{\rm m})$ WKB approximant,
$u_{w}^{\scriptscriptstyle{\rm WKB}}(x)$ has been computed to rather high (${\cal O}(\epsilon^{100})$) orders
in $\epsilon$.
Using then these results together with those of the core-expansion, Eq.\ \eqref{eq: core-exp2}, from \eqref{eq:um1} one finds for the parameter $x_0$ to ${\cal O}(\epsilon^{10})$
\begin{equation}\label{eq: x0}
x_0= -\,\frac{\alpha_{\rm m}}{4\gamma^4\epsilon}\left[1-9(\gamma\epsilon)^2
-113(\gamma\epsilon)^4-4987(\gamma\epsilon)^6
-\frac{663031}{2}(\gamma\epsilon)^8
-{\cal O}\left(\epsilon^{10}\right)\right]\,.
\end{equation}
Using \eqref{eq:um3} one may obtain the asymmetry in terms of $\alpha_{\rm m}$
\begin{equation}\label{eq: asymm1}
u_{-,3}(0)=\frac{\alpha_{\rm m}}{\epsilon ^3}\left[
1-5(\gamma\epsilon)^2-286(\gamma\epsilon)^4 -10422(\gamma\epsilon)^6-\frac{1227721}{2}(\gamma\epsilon)^8-{\cal O}(\epsilon^{10})
\right]\,.
\end{equation}

Inverting Eq.\ \eqref{eq: asymm1} in the limit $\epsilon\to0$, we find that $\alpha_{\rm m}$ is approximated to ${\cal O}(\epsilon^{2N+2})$
in terms of the asymmetry as
\begin{equation}\label{eq: asymm2}
\alpha_{\scriptscriptstyle\mathrm{WKB}}^{(N)}
=u_{-,3}(0)\epsilon ^3\left[
1+\sum_{n=1}^{N}a_n(\gamma\epsilon)^{2n}
\right]\,,
\end{equation}
where $a_n$ are positive rational numbers. The first four coefficients are
\begin{align}\label{eq: asymmcoeff}
 a_1=5 \,, \quad a_2=311 \,, \quad a_3=13407 \,,
 \quad a_4=\frac{1643903}{2} \,.
\end{align}
As further Supplemental Material to this paper we attach a Mathematica notebook file
\cite{Supp2},
which can be used to obtain the above results for $x_0$, $u_{-,3}(0)$ and $\alpha_{\scriptscriptstyle\mathrm{WKB}}^{(N)}$. The code can be used to calculate higher order generalizations up to order $\mathcal{O}(\epsilon^{100})$.

Before quantitatively comparing our results, \eqref{eq: asymm1}, \eqref{eq: asymm2}, with available numerical computations a few remarks are in order. As it should be clear from the previous part of this paper, the relations between the asymmetry and the amplitude of the wave tail deduced above, have been obtained in a perturbative framework, valid only for $\epsilon\ll1$. Furthermore, the wave tail is obtained from a
linearized approximation of Eq.\ \eqref{eq: statfKdV}, i.e.\ from Eq.\ \eqref{eq: uw}, in
form of a WKB asymptotic expansion in $\epsilon$. Therefore one expects that for a given value of $\epsilon\ll1$, Eqs.\ \eqref{eq: asymm1}, \eqref{eq: asymm2} can be used up to an $\epsilon$-dependent, ``optimal'' order of the WKB expansion.

\begin{figure}[!hbt]
 \centering
 \includegraphics[width=110mm]{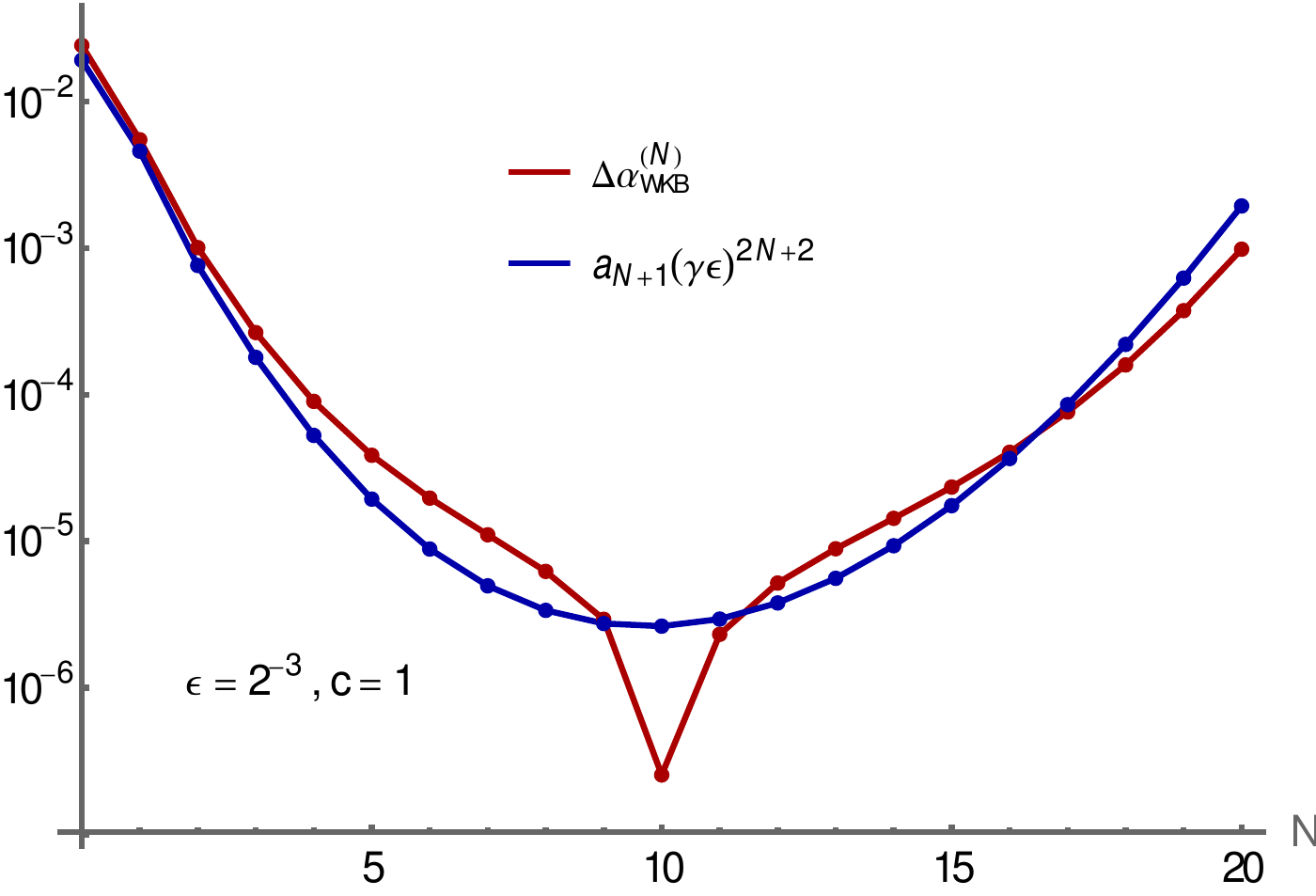}
\caption{\label{figalthd} The red points show the relative difference
$\Delta\alpha_{\scriptscriptstyle\mathrm{WKB}}^{(N)}$ for the parameter choice $\epsilon=2^{-3}$, $c=1$. The blue points give the relative contribution of the next term in \eqref{eq: asymm2},
$a_{N+1}(\gamma\epsilon)^{2N+2}$ which can be used to estimate the error of the approximation. Both take the smallest value for $N=10$, which is just one less than the optimal truncation order for the core expansion, $N_{\epsilon}=11$ given in Table \ref{tablerhoalp}.
}
\end{figure}

\begin{figure}[!hbt]
 \centering
 \includegraphics[width=110mm]{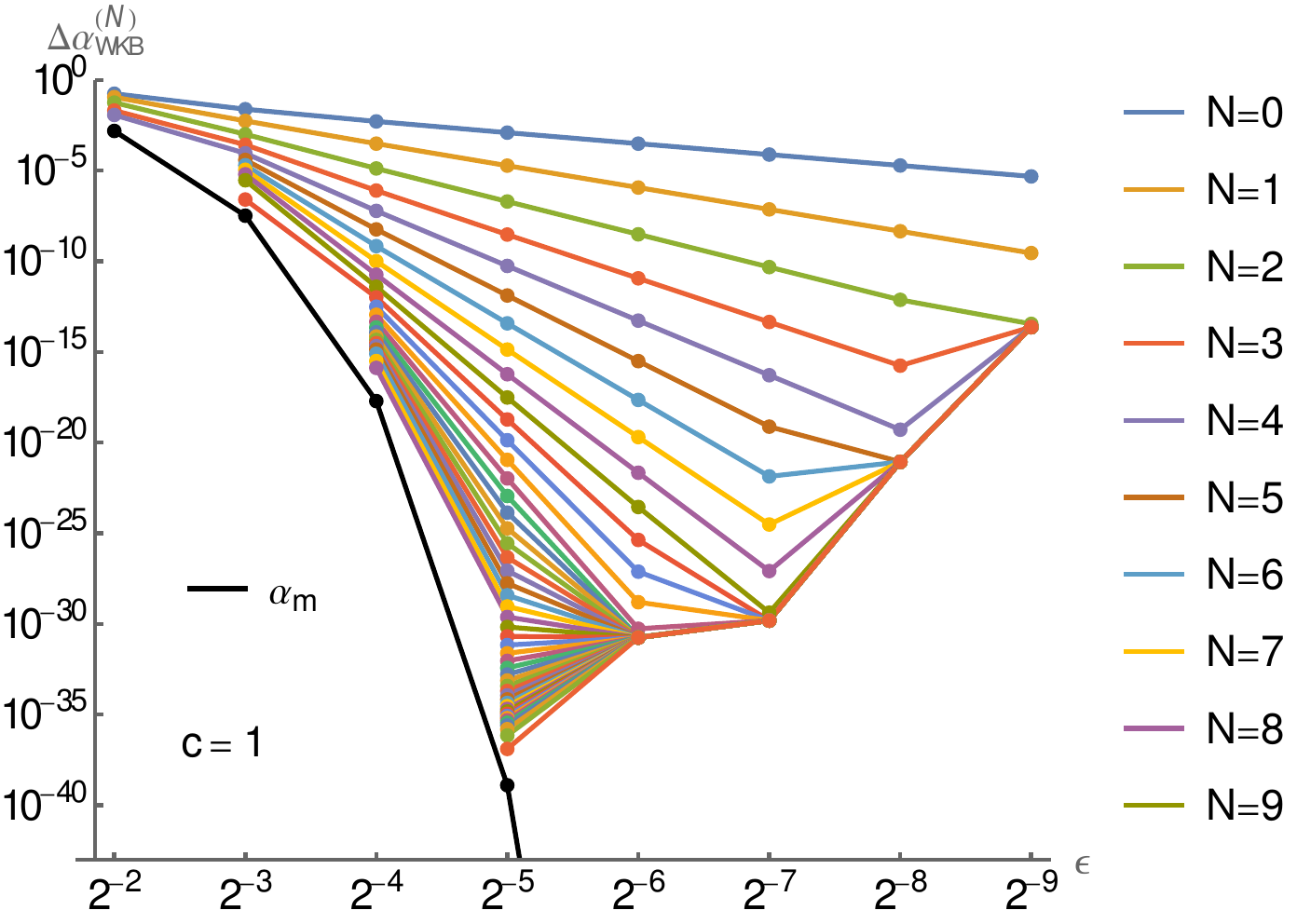}
\caption{\label{figerplot} For different values of $\epsilon$ we show how the relative difference $\Delta\alpha_{\scriptscriptstyle\mathrm{WKB}}^{(N)}$ changes as we increase the order of the approximation $N$. To make the plot more transparent we drop those points which give worse approximation than the lower order ones. For $\epsilon = 2^{-2}, 2^{-3}, 2^{-4}, 2^{-5}$ we stop at the optimal WKB truncation, $N=4,\, 10,\, 22,\, 48$, respectively. For $\epsilon < 2^{-5}$ our best spectral numerical values for $\alpha_{\rm m}$ are less precise than the higher order $\alpha_{\scriptscriptstyle\mathrm{WKB}}^{(N)}$ values, hence the points saturate at the numerical error. The relative error of the optimally truncated result approaches $\alpha_{\rm m}$, shown by the black points, consistently with the expectation that the error of both the numerical and WKB tail amplitude is of order $\alpha_{\rm m}^2$.}
\end{figure}

As we have seen at the end of Sec.~\ref{sec: comp}, for each specific choice of $\epsilon$ the central third derivative $u_{-,3}(0)$ can be calculated relatively easily and quickly up to hundreds of digits of precision. Hence it is a natural choice to use \eqref{eq: asymm2} to approximate the minimal tail amplitude $\alpha_{\rm m}$. The summation in the equation represents an asymptotic expansion, which has an optimal order of truncation.
For the choice $\epsilon=2^{-3}$ in Fig.~\ref{figalthd}
we plot the relative difference
\begin{equation}
\Delta\alpha_{\scriptscriptstyle\mathrm{WKB}}^{(N)}=\left|\left(\alpha_{\rm m}-\alpha_{\scriptscriptstyle\mathrm{WKB}}^{(N)}
\right)/\alpha_{\rm m}\right|
\end{equation}
of the numerically calculated $\alpha_{\rm m}$ and the approximation $\alpha_{\scriptscriptstyle\mathrm{WKB}}^{(N)}$ as a function of the number of terms $N$. Since the numerical value is much more precise in this case, $\Delta\alpha_{\scriptscriptstyle\mathrm{WKB}}^{(N)}$ represents the error of the $N$-th order approximation $\alpha_{\scriptscriptstyle\mathrm{WKB}}^{(N)}$. For all the $\epsilon$ values that we tested we have found that the best value of $N$ in $\alpha_{\scriptscriptstyle\mathrm{WKB}}^{(N)}$ agrees with $N_{\epsilon}$ or $N_{\epsilon}-1$, where $N_{\epsilon}$ is the optimal truncation order for the core expansion, listed in Table \ref{tablerhoalp}. Since in the small $\epsilon$ limit $N_{\epsilon}\sim 1.56/(\epsilon\sqrt{c})$, the error of the optimal order approximation by \eqref{eq: asymm2} decreases exponentially with decreasing $\epsilon$.
In Fig.~\ref{figerplot}
we show how the relative difference $\Delta\alpha_{\scriptscriptstyle\mathrm{WKB}}^{(N)}$ decreases for several values of $\epsilon$.

The above results clearly show that for small $\epsilon$ parameter values, compared to the spectral numerical code, it is more precise and efficient to use $u_{-,3}(0)$ and the WKB method to calculate the tail ampitude $\alpha_{\rm m}$ of the symmetric solution. The central derivative of the asymmetric solution $u_{-,3}(0)$ can be relatively easily calculated by the Hammersley-Mazzarino method presented in Sec~\ref{sec: comp}. for any $\epsilon$ which is larger than $2^{-15}$.
In Table \ref{tableamthd} we list the values of $\alpha_{\rm m}$ and $u_{-,3}(0)$ for smaller $\epsilon$ values than in Table \ref{tablerhoalp}.
\begin{table}[!hbtp]
 \centering
 \begin{tabular}{|c||c|c|}
  \hline
  $\epsilon$ & $\alpha_{\rm m}$  & $u_{-,3}(0)$ \\
  \hline
  \hline
  $2^{-10}$ & $4.95890409131\cdot10^{-1390}$ & $5.32457637656\cdot10^{-1381}$ \\
  \hline
  $2^{-11}$ & $1.50219375669\cdot10^{-2786}$ & $1.29037422688\cdot10^{-2776}$ \\
  \hline
  $2^{-12}$ & $3.43832141771\cdot10^{-5580}$ & $2.36279631071\cdot10^{-5569}$ \\
  \hline
  $2^{-13}$ & $4.49808545327\cdot10^{-11168}$ & $2.47284858324\cdot10^{-11156}$ \\
  \hline
  $2^{-14}$ & $1.92344734444\cdot10^{-22344}$ & $8.45941084312\cdot10^{-22332}$ \\
  \hline
  $2^{-15}$ & $8.79023705549\cdot10^{-44698}$ & $3.09278970949\cdot10^{-44684}$ \\
  \hline
 \end{tabular}
 \caption{For the $\epsilon$ values in this table the central asymmetry $u_{-,3}(0)$ of the solution $u_{-}$ is used to calculate the minimal tail amplitude $\alpha_{\rm m}$ of the symmetric solution.}
 \label{tableamthd}
\end{table}
In these cases $\alpha_{\rm m}$ has been calculated from $u_{-,3}(0)$ using appropriately many terms in \eqref{eq: asymm2}. In Table \ref{tableamthd} we give ``only'' 12 decimal digits, but we have calculated these results to several hundred digits.

\begin{figure}[!hbt]
 \centering
 \includegraphics[width=110mm]{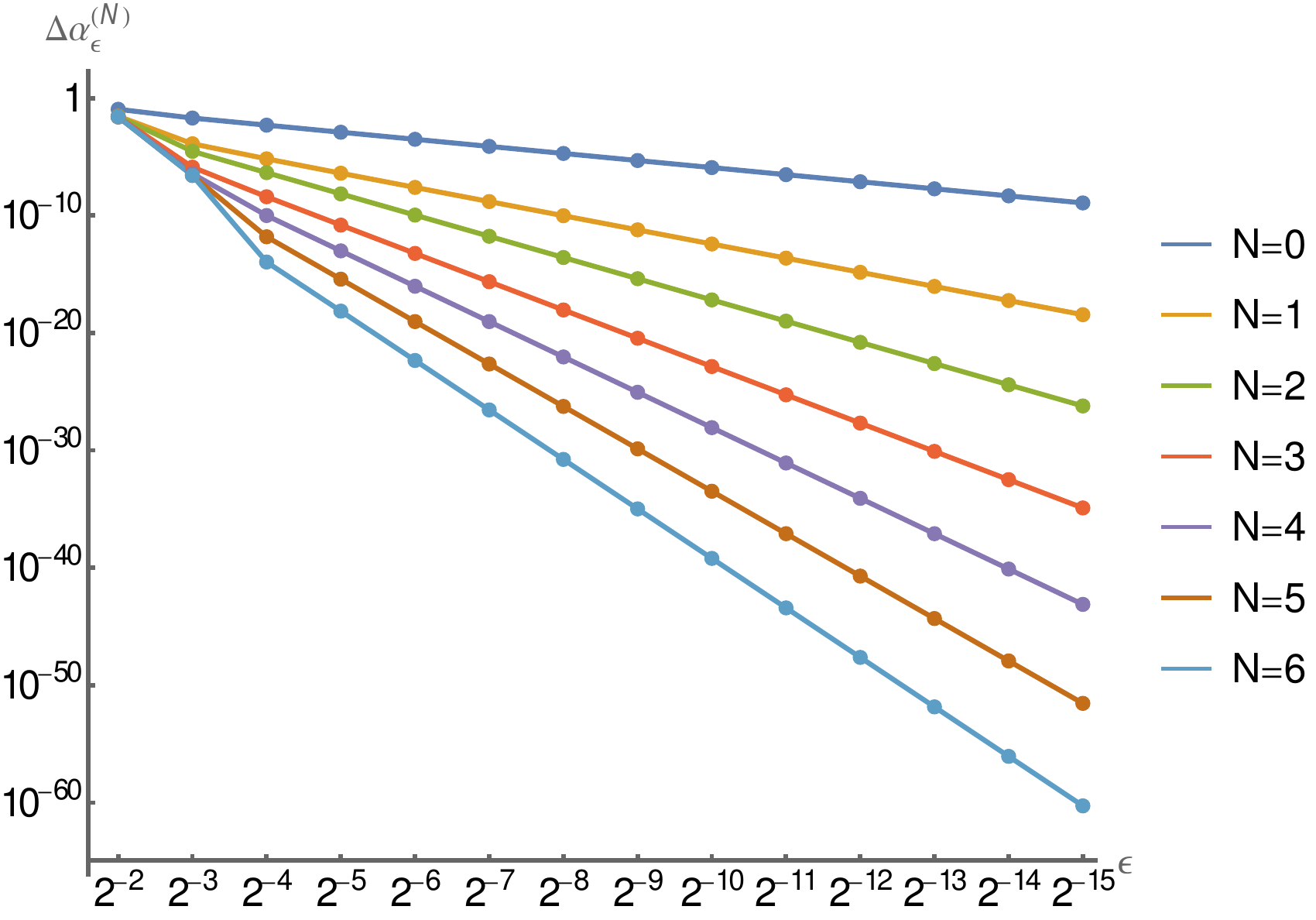}
\caption{\label{figexpcmp} Relative difference of the $N$-th order $\epsilon$ expansion result $\alpha_{\rm m}^{(N)}$ for the tail amplitude and a more precise high order WKB result, $\alpha_{\scriptscriptstyle\mathrm{WKB}}^{(N_\epsilon)}$.}
\end{figure}

It is instructive to compare the tail amplitude, $\alpha_{\scriptscriptstyle\mathrm{WKB}}^{(N_\epsilon)}$, calculated by the optimal order WKB method to the $\epsilon$ expansion results of $\alpha_{\rm m}^{(N)}$ in Eq.\ \eqref{eq: alpha}. In Fig.~\ref{figexpcmp}
we depict the relative difference
$\Delta\alpha_\epsilon^{(N)}=\left|\left(
\alpha_{\scriptscriptstyle\mathrm{WKB}}^{(N_\epsilon)}
-\alpha_{\rm m}^{(N)}\right)
/\alpha_{\scriptscriptstyle\mathrm{WKB}}^{(N_\epsilon)}
\right|$
for a large range of $\epsilon$.
The compact formula of the ${\cal{O}}(\epsilon^{12})$ result for $\alpha_{\rm m}^{(N)}$ given by Eq.\ \eqref{eq: alpha} is simple to evaluate. The WKB expansion is known to much higher orders $(\gtrsim{\cal{O}}(\epsilon^{200})$),
implying it being much more precise. The WKB result is, however, prohibitively complicated to be displayed to such high orders. We note, that for $\epsilon<2^{-4}$ $\alpha_{\rm m}^{(6)}$ yields more than $16$ significant decimal digits, which is good enough for most cases.

The $\epsilon$ expansion result \eqref{eq: alpha} for the symmetric solution can also be used together with the WKB transformation \eqref{eq: asymm1} to calculate the central third derivative of the asymmetric solution. Since the coefficients $\xi_n$ are known numerically, we give only numerical coefficients here,
\begin{align}\label{eq: um3eps}
 u_{-,3}(0)&=\frac{\mathcal{K}}{\epsilon^5}
 {\rm e}^{-\tfrac{\pi k}{2\gamma\epsilon}}
  \bigl[ 1-10(\gamma\epsilon)^2-267.544068194(\gamma\epsilon)^4
  -9433.69349852(\gamma\epsilon)^6 \notag\\
  &-561740.239659(\gamma\epsilon)^8-45224879.2541(\gamma\epsilon)^{10}
  -4611707502.31(\gamma\epsilon)^{12}-\ldots
  \bigr]\,,
\end{align}
where $\mathcal{K}\approx19.96894735876096\!\cdot\!\pi$.
It is now easy to obtain the $\tilde\epsilon$ expansion of the asymmetry, $Y_3=y_{,3}(0)$, to leading order in $\alpha_{\rm m}$, using Eqs.\ \eqref{eq: 3d}, \eqref{eq: k}, \eqref{eq: kgamma} alternatively \eqref{eq: cgamma} :
 \begin{align}\label{eq: ym3eps}
 Y_3&=\frac{3\mathcal{K}}{\tilde\epsilon^{5/2}}
 {\rm e}^{-\tfrac{\pi k^2}{\sqrt{\tilde\epsilon}}}
  \biggl[ 1-\frac{5}{2}\tilde\epsilon-14.2215042621\tilde\epsilon^2
  -118.958452390\,\tilde\epsilon^3 \notag\\
  &-1823.20094973\,\tilde\epsilon^4-36515.2419905\,\tilde\epsilon^5
  -932270.777827\tilde\epsilon^6-\ldots
  \biggr]\,,
\end{align}
In Eq.\ \eqref{eq: ym3eps} $k^2$ in the exponential has not been expanded in $\tilde\epsilon$ to avoid the appearance of half integer powers of $\tilde\epsilon$, since
\begin{equation}
 {\rm e}^{-\tfrac{\pi k^2}{\sqrt{\tilde\epsilon}}}=
 {\rm e}^{-\tfrac{\pi}{\sqrt{\tilde\epsilon}}}
 \left[1-\pi\tilde\epsilon^{1/2}+\frac{\pi^2}{2}\tilde\epsilon
 -\frac{\pi}{6}(\pi^2-6)\tilde\epsilon^{3/2}+\ldots
 \right] .
\end{equation}
At this point we should like to point out, that the leading order term for $\tilde\epsilon\to 0$ of the asymmetry,
is in complete agreement with a previous result of Byatt-Smith in Ref.\ \cite{Byatt-Smith91}. In Ref.\ \cite{Byatt-Smith91} it has been used to provide yet another proof of the non-existence of a globally regular decaying solution on $\mathbb{R}$.

The perturbative results in Eqs.\ \eqref{eq: um3eps} and \eqref{eq: ym3eps} are given in terms of asymptotic series. E.g.\ for $\tilde\epsilon=2^{-4}$ the optimal truncation corresponds to keeping the first four terms in \eqref{eq: ym3eps}, up to order $\tilde\epsilon^3$.
This way one obtains $y_{,3}(0)=0.243033$, which is correct to 4 significant digits,
the result up to 7 significant digits being $Y_3=0.243445$. For
$\tilde\epsilon\leq 2^{-8}$ the seven terms in \eqref{eq: ym3eps} give at least six significant digits, hence we obtain the values of $Y_3$ listed in the first 16 lines of Table 1.\ of Ref.\ \cite{HM}. For $\tilde\epsilon\leq 2^{-20}$ the
the leading order approximation is already accurate to six decimal digits. The method presented in Sec.~\ref{sec: comp} can be used for the calculation of the third derivative for essentially any $\tilde\epsilon$ up to hundreds of digits of precision. However, for $\tilde\epsilon\leq 2^{-6}$ it is much easier to use the simple expression in Eq.~\eqref{eq: ym3eps} to obtain a suitably precise value.

\section{Conclusions}\label{sec:conc}
In the limit $\epsilon\to0$, a connection has been established between the amplitude of the minimal wave tail of weakly delocalized solitons of the fKdV equation and the asymmetry of (smooth) solutions on $\mathbb{R}^+$ tending to zero for $x\to\infty$, constructed in Ref.\ \cite{HM}.
The asymmetry of the decaying solutions of Ref.\ \cite{HM} is an exact, albeit somewhat implicit result.
We have simplified a bit the numerical calculation of the asymmetry, and on a small laptop computer it is easily
computed to many significant digits, as it can be verified from our Supplemental Material.
On the basis of our high order perturbative results for the tail amplitude, a small $\epsilon$ expansion of the asymmetry
has been derived and verified numerically.
Our results make the computation of the tail amplitudes possible for arbitrarily small values of $\epsilon$ to extremely high precision.
Exploting this correspondence in combination with high order perturbative calculations and precise numerical methods, we have been able to compute for such small values as $\epsilon\leqslant 2^{-15}$, in which case the amplitude of the minimal wave tail is $\sim10^{-44698}$. Somewhat surprisingly our method works quite well up to as ``large'' values of $\epsilon$ as $2^{-3/2}\approx0.353553$.

\section*{Acknowledgements}

The research of G. Fodor has been supported in part by NKFIH OTKA Grants No. K 138277 and K 142423.

\bibliography{fkdv}

\end{document}